\documentclass[preprintnumbers,superscriptaddress,amsmath,amssymb,prd,preprint,nofootinbib]{revtex4-2}
\usepackage{graphicx}
\usepackage{amsfonts}
\usepackage[colorlinks,linkcolor=red,anchorcolor=blue,citecolor=green]{hyperref}
\usepackage{subfigure}
\usepackage{float}
\usepackage{enumerate}
\newcommand\diff{\mathrm{d}}
\newcommand\Diff{\mathcal{D}}

\newcommand\e{\mathrm{e}}

\begin{document}
\title{
Static plane symmetric solutions in $f(Q)$ gravity
}
\author{Jun-Qin~Long}
\author{Rui-Hui Lin}
\email[]{linrh@shnu.edu.cn}
\author{Xiang-Hua Zhai}
\email[]{zhaixh@shnu.edu.cn}
\affiliation{Division of Mathematics and Theoretical Physics, Shanghai Normal University, 100 Guilin Road, Shanghai 200234, China}

\begin{abstract}
We systematically investigate static plane symmetric configurations in $f(Q)$ gravity.
For vacuum regions, we discuss the constancy of the nonmetricity scalar $Q$ 
and derive general vacuum solutions, 
which correspond effectively to Taub-(anti) de Sitter spacetimes 
with a cosmological constant determined by the specific $f(Q)$ model. 
By matching a singular thin shell source to the vacuum solutions, 
we relate the shell's energy density and pressure to the integration constants of the exterior geometry.
We also examine a finite-thickness slab as another matter source supporting the vacuum solution. 
Through numerical analysis of a quadratic model $f(Q)=Q+\alpha Q^2$ with isotropic matter, 
we show that the maximum pressure inside the slab generally does not coincide with the geometric center. 
Moreover, a negative $\alpha$ with larger magnitude leads to higher internal pressure and a thicker slab, 
while models with positive $\alpha$ are incompatible with a self-gravitating slab of positive pressure.
\end{abstract}
\maketitle

\newpage
\section{Introduction}
\label{intro}
The standard formulation of general relativity (GR) is founded upon Riemannian geometry,
adopting the Levi–Civita connection as the affine connection on the spacetime manifold, 
which by definition is torsion-free and metric-compatible.
However, a manifold admits multiple possible choices of affine connection, 
and distinct connections can yield different, yet physically equivalent, descriptions of gravity \cite{BeltranJimenez:2019esp,PhysRevD.101.024053}, 
potentially offering complementary theoretical insights. 
In GR, the Levi–Civita connection implies that, apart from curvature $R$, 
the two other fundamental geometric quantities, nonmetricity $Q$ and torsion $T$, are both zero. 
By relaxing these constraints, gravitational theories can be formulated within a non-Riemannian geometric framework, 
where curvature, torsion, and nonmetricity may all be nonzero. 
In particular, imposing vanishing curvature and nonmetricity while allowing nonzero torsion leads to the teleparallel equivalent of general relativity (TEGR) \cite{Maluf2013,Aldrovandi:2013wha}. 
As a third alternative, a flat and torsion-free geometry with nonvanishing nonmetricity gives rise to the symmetric teleparallel formulation of GR (STGR) \cite{Nester:1998mp,Adak:2005cd,Mol:2014ooa,PhysRevD.98.044048}.

Despite the empirical success of GR, long-standing theoretical and observational challenges, 
such as the nature of dark content and the difficulties in quantizing gravity, remain unresolved, 
motivating extensive research into modified theories of gravity. 
The equivalent reformulations of GR mentioned above inherit these issues.
Yet, the geometric alternatives open new avenues for modifying GR. 
One prominent strategy to modification is to generalize the action by replacing its foundational scalar with an arbitrary function. 
This approach is exemplified in the standard GR formulation by $f(R)$ gravity
(see, e.g., Refs. \cite{Sotiriou:2008rp,DeFelice:2010aj,Nojiri:2010wj,Capozziello:2011et} for comprehensive reviews). 
In direct analogy, one may generalize TEGR by constructing $f(T)$ gravity, in which the torsion scalar $T$ is promoted to a function $f(T)$ 
(see, e.g., Refs. \cite{Cai:2015emx,Nojiri:2017ncd,Lin:2021ijx,Yang:2022efz}).

Following a similar rationale that attributes the dark sector to geometric properties of spacetime,
one may start from STGR and construct the $f(Q)$ gravity.
One of the primary benefits of these kinds of modification, like the $f(T)$ schemes,
is that they feature in second order field equations instead of fourth order ones in $f(R)$ gravity.
Over recent decades, $f(Q)$ gravity has garnered considerable attention,
especially in the context of cosmology including various aspects
(see, e.g., Refs. 
\cite{BeltranJimenez:2019tme,Lazkoz:2019sjl,Lu:2019hra,Barros:2020bgg,Dimakis:2022rkd,Khyllep:2022spx,Subramaniam:2023okn,Bajardi:2023vcc,Heisenberg:2023lru,Guzman:2024cwa}). 
Beyond cosmology, the framework has also been actively applied to black hole physics 
\cite{DAmbrosio:2021zpm},
stellar structure solutions \cite{Lin:2021uqa,Wang:2021zaz,Calza:2022mwt}.

In parallel to the extensive investigations into $f(Q)$ gravity across various fronts,
a complementary and fundamental strand of research within this framework focuses on deriving solutions with high symmetry,
which remains a cornerstone of gravitational physics.
Such solutions may provide critical insights into a theory's fundamental structure and novel phenomenological implications.
Theoretical groundwork addressing essential issues such as
connection gauge fixing and symmetrical degrees of freedom has laid a basis for this pursuit
\cite{Lin:2021uqa,Bahamonde:2022zgj,Hohmann:2024phz}.
In particular, spherical symmetric configurations have been explored,
providing a primary testing ground \cite{DAmbrosio:2021zpm,Wang:2021zaz,Zhao:2021zab}.
In this context, plane symmetry offers another ideal arena for such explorations.
Spacetimes endowed with this symmetry provide a rich yet tractable framework for understanding gravitational dynamics in settings beyond spherical symmetry.
Moreover, plane symmetric gravitational backgrounds may also be of significant interest in quantum theory or string theory, 
where they may serve as simplified models for studying domain walls and the formation of gravitational bound states \cite{Ipser:1984db,PhysRevD.74.125014,Lanosa:2023mox}.

The static plane symmetric vacuum solution in GR, known as the Taub solution, was established in 1951 \cite{Taub:1951ez},
with its gravitational properties subsequently elaborated \cite{Aichelburg:1970,Bedran:1997su}.
Beyond vacuum scenarios, a significant body of research has been dedicated to plane symmetric systems with material sources,
involving perfect fluids 
\cite{Groen:1992sm,Jensen:1994zf,Bedran:1997su,GamboaSaravi:2007se}
and, more specifically, finite-thickness slabs of matter
\cite{Bedran:1997su,daSilva:1997uf,GamboaSaravi:2009sw,Fulling:2015uva,Acuna:2015tsa,Kamenshchik:2020div}.
This line of inquiry was recently extended to higher dimensions \cite{Avagyan:2024oia}.
Furthermore, a new class of plane symmetric solutions was found and analyzed by Zhang et al \cite{Zhang:2008ss,Zhang:2009hx},
which has also been explored in higher-dimensional spacetimes \cite{Zhang:2009hz,Zhang:2011zzq}.
The pursuit of plane symmetric solutions has naturally been extended to various modified gravity theories.
This includes comprehensive studies within the frameworks of $f(R)$ \cite{Sharif:2009uc,Yavari:2013fga,Oz:2021jnw,Agrawal:2017wvt,Mehmood:2022zcg,Singh:2023mgr,Dalal:2024qkm}
and $f(T)$ gravities \cite{Sharif:2012fst,RODRIGUES_2013}.

In this work, we intend to study the plane symmetric configurations in $f(Q)$ gravity.
The paper is organized as follows.
In Sec. \ref{fQreview}, we briefly review the nonmetricity description of gravity
and set the static plane symmetric configuration for $f(Q)$ gravity.
The vacuum solutions and thin shell source are discussed in Sec. \ref{exact}.
Section \ref{slab} is dedicated to the analysis of matter slabs of finite thickness.
We summarize our work in Sec \ref{conclusion}.
Throughout the paper, we will be working in the natural unit where $8\pi G=c=1$.

\section{Static symmetric in $f(Q)$ gravity}
\label{fQreview}
\subsection{Nonmetricity framework of GR}
Let the spacetime manifold $\mathcal{M}$ be a parallelizable and differentiable metric space,
its geometry is generally characterized by a non-degenerate Lorentzian metric $g$ 
and a linear connection $\Gamma$.
The full connection 1-form $\Gamma^a_{\;\:b}$, 
associated with a covariant derivative $\Diff(\Gamma)$,
admits a unique decomposition\cite{Lin:2021uqa}
\begin{equation}
    \label{connection}
    \Gamma^a_{\;\:b}=\omega^a_{\;\:b}+K^a_{\;\:b}+L^a_{\;\:b},
\end{equation}
where $\omega^a_{\;\:b}$ is the Levi-Civita connection 1-form with
\begin{equation}
    \label{covderw}
    \Diff(\omega) \vartheta^a=d \vartheta^a+\omega^a_{\;\:b}\wedge \vartheta^b=0,
\end{equation}
$K^a_{\;\:b}$ denotes the contorsion 1-form
and $L^a_{\;\:b}$ the disformation 1-form.
Here, $d$ represents the exterior derivative,
and $\{\vartheta^a\}$ is the dual coframe corresponding to a general frame $\{e_a\}$
satisfying $\vartheta^b(e_a)=\delta^b_a$.
We will be working in the torsion-free setting that $K^a_{\;\:b}=0$.
Then the nonmetricity 1-form associated with this connection is given by
\begin{equation}
    \label{nonmetric0}
    \begin{split}
        Q_{ab}=&\frac12\Diff(\Gamma)\eta_{ab}=\Gamma_{(ab)}=-L^c_{\:\;b}\eta_{ac}-L^c_{\:\;a}\eta_{cb},
    \end{split}
\end{equation}
where $\Diff(\omega)\eta_{ab}=0$ is applied.
In component form,
\begin{equation}
    \label{Qcom}
    Q_{\alpha\beta\gamma}=\nabla_\alpha g_{\beta\gamma}=-L^\rho_{\alpha\beta}g_{\rho\gamma}-L^\rho_{\alpha\gamma}g_{\rho\beta},
\end{equation}
or equivalently,
\begin{equation}
    \label{Qcom1}
    L^\alpha_{\beta\gamma}=\frac12Q^\alpha_{\beta\gamma}-Q_{(\beta\gamma)}^{\quad\;\alpha}.
\end{equation}
The curvature 2-form are given by
\begin{equation}
    \label{curvature0}
    R^a_{\;\:b}(\Gamma)=\Diff(\Gamma)\Gamma^a_{\;\:b}=d\Gamma^a_{\;\:b}+\Gamma^a_{\;\:c}\wedge\Gamma^c_{\;\:b}.
\end{equation}
When $\Gamma^a_{\;\:b}=\omega^a_{\;\:b}$, Eq. \eqref{curvature0} becomes the Riemannian curvature,
\begin{equation}
    \label{curvature1}
    R^a_{\;\:b}(\omega)=d\omega^a_{\;\:b}+\omega^a_{\;\:c}\wedge\omega^c_{\;\:b},
\end{equation}
and, thus,
\begin{equation}
    \label{curvature2}
    R^a_{\;\:b}(\Gamma)=R^a_{\;\:b}(\omega)+\Diff(\omega)L^a_{\;\:b}+L^a_{\;\:c}\wedge L^c_{\;\:b}.
\end{equation}
Let $h^{ab\cdots}=*(\vartheta^a\wedge\vartheta^b\wedge\cdots)$, 
then the curvature 4-form becomes
\begin{equation}
    \label{curvaturescalar0}
    \begin{split}
        R(\Gamma)=&R^a_{\;\:b}\wedge h^b_{\;\:a}\\
        =&R(\omega)+\Diff(\omega)L^a_{\;\:b}\wedge h^b_{\;\:a}+L^a_{\;\:c}\wedge L^c_{\;\:b}\wedge h^b_{\;\:a}\\
        =&R(\omega)+d(A^a_{\;\:b}\wedge h^b_{\;\:a})-L^a_{\;\:b}\wedge\Diff(\omega)h^b_{\;\:a}+L^a_{\;\:c}\wedge L^c_{\;\:b}\wedge h^b_{\;\:a}.
    \end{split}
\end{equation}
Note that $R(\omega)$ constitutes the Hilbert-Einstein Lagrangian 4-form of the standard formulation of GR.
To describe gravitation purely in terms of nonmetricity,
i.e., the connection is also curvature-free and $R^a_{\;\:b}(\Gamma)=0$
\cite{Nester:1998mp,Adak:2005cd,Adak:2008gd,PhysRevD.98.044048}, we have
\begin{equation}
    \label{HELag}
    \begin{split}
        \mathcal L_\text{EH}=&\frac12R(\omega)\\
        =&-d\left(\frac12L^a_{\;\:b}\wedge h^b_{\;\:a}\right)+\frac12L^a_{\;\:c}\wedge L^c_{\;\:b}\wedge h^b_{\;\:a},
    \end{split}
\end{equation}
where $\Diff(\omega)h^b_{\;\:a}=0$ has been taken into account.
Discarding the exact form and introducing the nonmetricity 4-form $\mathcal Q=L^a_{\;\:c}\wedge L^c_{\;\:b}\wedge h^b_{\;\:a}$,
one obtains the Lagrangian 4-form of STGR
\begin{equation}
    \label{STGRLag}
    \mathcal L_\text{STGR}=\frac12L^a_{\;\:c}\wedge L^c_{\;\:b}\wedge h^b_{\;\:a}=\frac12\mathcal Q.
\end{equation}
In component form, this reads
\begin{equation}
    \label{STGRLag1}
    \mathcal L_\text{STGR}=-\frac{\sqrt{-g}}{2}g^{\mu\nu}\left( L^\alpha_{\beta\nu}L^\beta_{\mu\alpha}-L^\beta_{\alpha\beta}L^\alpha_{\mu\nu} \right)\diff^4 x=\frac{\sqrt{-g}}{2}Q\diff^4x,
\end{equation}
where $Q\equiv -g^{\mu\nu}\left( L^\alpha_{\beta\nu}L^\beta_{\mu\alpha}-L^\beta_{\alpha\beta}L^\alpha_{\mu\nu} \right)$
is the nonmetricity scalar.

\subsection{$f(Q)$ gravity}
In $f(Q)$ gravity, the nonmetricity scalar $Q$ in Lagrangian \eqref{STGRLag1}
is generally replaced by an arbitrary function $f(Q)$,
leading to the gravitational action
\begin{equation}
    \label{action}
    \mathcal{S}=\frac1{2}\int\sqrt{-g}f(Q)\diff^4x.
\end{equation}
The corresponding field equation then takes the form
\begin{equation}
    \label{eomfQ4}
    -\frac2{\sqrt{-g}}\nabla_\alpha \left( \sqrt{-g}f_QP^\alpha_{\;\mu\nu} \right)+f_Q \left( P_\nu^{\;\alpha\beta}Q_{\mu\alpha\beta}-2P^{\alpha\beta}_{\;\:\;\:\mu}Q_{\alpha\beta\nu} \right)+\frac12g_{\mu\nu}f= \mathcal{T}_{\mu\nu}.
\end{equation}
where $f_Q=\diff f(Q)/\diff Q$, $f_{QQ}=\diff^2 f(Q)/\diff Q^2$,
$\nabla_\alpha$ denotes the component of the covariant derivative that associates to $\Gamma$,
and the nonmetricity conjugate $P^\alpha_{\;\beta\gamma}$ is given by
\begin{equation}
    \label{Pdef}
    P^\alpha_{\;\beta\gamma}=-\frac12L^\alpha_{\beta\gamma}+\frac{1}{4}\left( Q^\alpha-\tilde{Q}^\alpha \right)g_{\beta\gamma}-\frac14\delta^\alpha_{(\beta}Q_{\gamma)},
\end{equation}
with $Q_\alpha=Q_{\alpha\beta}^{\quad\beta}$ and $\tilde{Q}_\alpha=Q^\beta_{\;\alpha\beta}$.
With the component form of the Riemannian curvature $R^a_{\;\:b}(\omega)$
and its contractions, Eq.\eqref{eomfQ4} can also be expressed in the form
\begin{equation}
    \label{eomfQ3}
        2f_{QQ}P^\alpha_{\;\:\mu\nu}\partial_\alpha Q+\frac12 g_{\mu\nu}\left( f-f_QQ \right)+f_Q G_{\mu\nu}= \mathcal T_{\mu\nu}.
\end{equation}
It is straightforward to verify that the Einstein equation is recovered when $f(Q)=Q$.

\subsection{static plane symmetry}
The most general static metric possessing plane symmetry can be written in the form
\begin{equation}
    ds^2=e^{2u(z)}dt^2-dz^2-e^{2v(z)}(dx^2+dy^2),
\end{equation}
with the corresponding Killing vectors given by
\begin{equation}
\begin{split}
\xi_{(t)}=\begin{pmatrix}1 \\0 \\0 \\0\end{pmatrix},\;
\xi_{(x)}=\begin{pmatrix}0 \\1 \\0 \\0\end{pmatrix},\;
\xi_{(y)}=\begin{pmatrix}0 \\0 \\1 \\0\end{pmatrix},\;
\xi_{(\theta)}=\begin{pmatrix}0  \\-y \\x  \\0\end{pmatrix}.
\end{split}
\end{equation}
A connection preserving this symmetry should satisfy
\begin{equation}
    \label{lieD}
    \mathcal{L}_{\xi_{(i)}} \Gamma^\lambda_{\;\mu\nu}
    = \frac{\partial \Gamma^\lambda_{\;\mu\nu}}{\partial x^\rho} \xi_{(i)}^\rho
    + \frac{\partial\xi_{(i)}^\rho }{\partial x^\mu} \Gamma^\lambda_{\;\rho\nu}
    + \frac{\partial \xi_{(i)}^\rho }{\partial x^\nu} \Gamma^\lambda_{\;\mu\rho}
    - \frac{\partial \xi_{(i)}^\lambda }{\partial x^\rho} \Gamma^\rho_{\;\mu\nu}
    + \frac{\partial^2 \xi_{(i)}^\lambda }{\partial x^\mu \partial x^\nu} =0
\end{equation}
for $i=t,x,y,\theta$,
which impose additional constraints beyond the conditions of vanishing curvature and torsion.
Although, as seen in the cosmological and spherical cases \cite{Zhao:2021zab,Dimakis:2022rkd,Subramaniam:2023okn},
solving the full combined system is intricate,
it is straightforward to verify that $\Gamma^\lambda_{\;\mu\nu}=0$
represents a valid set of solution compatible with the current symmetry.
Therefore, we adopt this vanishing connection for simplicity.
Consequently, the nonvanishing components of the tensors
$Q_{\alpha\beta\gamma}$ and $L^\alpha_{\;\;\beta \gamma}$ are
\begin{equation}
    \label{Pnonmetricity}
    \begin{split}
        &Q_{ztt}=2e^{2u}u',\\
        &Q_{zxx}=-2e^{2v}v',\\
        &Q_{zyy}=-2e^{2v}v',
    \end{split}
\end{equation}
and
\begin{equation}
    \label{Pdeformation}
    \begin{split}
        &L^t_{\;\;tz}=L^t_{\;\;zt}=-u',\\
        &L^x_{\;\;xz}=L^x_{\;\;zx}=L^y_{\;\;yz}=L^y_{\;\;zy}=-v',\\
        &L^z_{\;\;tt}=-e^{2u}u', L^z_{\;\;xx}=L^z_{\;\;yy}=e^{2v}v',
    \end{split}
\end{equation}
respectively, where the prime symbol $'$ indicates derivative with respect to the perpendicular direction $z$.
Consequently, the nonmetricity scalar $Q$ reads
\begin{equation}
    \label{PQ}
    Q=2v'(2u'+v').
\end{equation}
It follows that the field equations are
\begin{equation}
    \label{Peom}
    \begin{split}
          \mathcal{T}_{tt} 
    =& \frac{1}{2} e^{2u} \left\{ f(Q) - 4 f_{Q} (v''+2v'^2+u' v')\right.\\
    &\left. - 16f_{QQ}(u'' v'^2 +v'' v'^2 +u' v' v'') \right\}, \\
      \mathcal{T}_{xx} =   \mathcal{T}_{yy} 
    =&\frac{1}{2} e^{2v} \left\{ -f(Q) + 2 f_{Q} \left[u'' +  v'' + 2 v'^2   +  u'^2 + 3u' v'\right] \right.\\
        &\left.+ 8 f_{QQ}\left[v'^2 (u'' +v'')+ u'^2 v'' + u' v' (u'' +2 v'')\right]  \right\},\\
        \mathcal{T}_{zz} 
    =& -\frac{f(Q)}{2} + 4f_{Q} u' v' + 2f_{Q} v'^2 .
    \end{split}
\end{equation}
Having established the field equations, we now proceed to seek their solutions.

\section{Vacuum regions and thin shell}
\label{exact}
\subsection{External vacuum solutions}
\label{vac}
In this subsection, we focus on the vacuum scenario with $\mathcal{T}_{\mu\nu}=0$.
From Eq. \eqref{PQ}, 
we observe that for the vacuum case the $zz$ component of Eq. \eqref{Peom} becomes
\begin{equation}
    \label{PAlgebeq}
    -\frac{f(Q)}{2} + f_{Q} Q =0.
\end{equation}
Given a specific form of the function $f(Q)$,
this expression can be viewed as an algebraic equation for the scalar $Q$.
This implies the existence of algebraic solutions where $Q$ is fixed by the explicit functional form of $f$.
Under the principle that the gravitational theory, 
and hence the functional form $f$ in its Lagrangian, 
are fundamental and universal,
the value of $Q$ derived in this manner must be invariant across spacetime.
It follows that in vacuum, $\partial_\mu Q=0$, and thus $Q = Q_0$ is a constant.
It is worth noting that, 
unlike in $f(R)$\cite{Sharif:2009uc} or $f(T)$\cite{RODRIGUES_2013} gravity 
where the curvature scalar $R$ or torsion scalar $T$ is sometimes artificially set to be constant, 
the constancy of the nonmetricity scalar $Q$ in this context arises 
as a direct consequence of the field equations.

In this scenario, the $tt$ component of Eq. \eqref{Peom} reduces to
\begin{equation}
    \label{finaleq1}
    2v''+3v'^2=\frac{Q_0}{2}.
\end{equation}
Together with Eq. \eqref{PQ},
these two equations close the system for the metric functions $u(z)$ and $v(z)$.
For $Q_0>0$, the solution can be written as
\begin{equation}
    \label{plusQ0}
    \begin{split}
        u(z)&=-\frac{1}{3} \ln  \left\lvert \cosh  \sqrt{\frac{3Q_0}{8}}(z+A)\right\rvert  + \ln \left\lvert \sinh \sqrt{\frac{3Q_0}{8}}(z+A)\right\rvert +B, \\
        v(z)&=\frac{2}{3} \ln  \left\lvert \cosh  \sqrt{\frac{3Q_0}{8}}(z+A)\right\rvert+C,
    \end{split}
\end{equation}
where $A$, $B$ and $C$ are integration constants.
Similarly, for $Q_0<0$, the corresponding solution takes the form
\begin{equation}
    \label{minusQ0}
    \begin{split}
        u(z)&=-\frac{1}{3} \ln  \left\lvert \cos  \sqrt{\frac{3\left\lvert Q_0\right\rvert }{8}}(z+A)\right\rvert  + \ln \left\lvert \sin \sqrt{\frac{3\left\lvert Q_0\right\rvert }{8}}(z+A)\right\rvert +B,\\
        v(z)&=\frac{2}{3} \ln  \left\lvert \cos  \sqrt{\frac{3\left\lvert Q_0\right\rvert }{8}}(z+A)\right\rvert+C.
    \end{split}
\end{equation}

These solutions in fact correspond to the Taub-(anti) de Sitter solutions \cite{Stephani:2003tm}
with the cosmological constant given by $\Lambda =-\frac{Q_0}{2}$.
This identification is also supported by the fact that 
the field equation \eqref{eomfQ3} reduces to the Einstein equations with a cosmological constant
\begin{equation}
    \label{ttcomeq}
    \frac{1}{2}g_{\mu\nu}Q_0 + G_{\mu\nu}=0,
\end{equation}
when $Q_0$ is constant.

However, the standard Taub solution cannot be recovered by simply setting $Q_0=0$ in Eq. \eqref{plusQ0} or \eqref{minusQ0},
even though Eq.\eqref{ttcomeq} reduces formally to the Einstein equation in this limit.
The underlying reason is that the signs of $Q_0$,
may that be positive, zero, or negative,
define three structurally distinct branches
of the differential equations governing $u(z)$ and $v(z)$.
Therefore, moving between these branches is not a matter of continuous parameter variation 
but involves a qualitative change in the system description.

\subsection{Thin shell matter source}
In this subsection, we consider a static ideal fluid with plane symmetry, confined to the plane $z=0$,
which may serve as the source of the external vacuum obtained above.
For such matter,
the energy-momentum tensor is given by
\begin{equation}
    \label{singularT}
\mathcal{T}^{\mu}_{\;\;\;\;\nu}=
\begin{pmatrix}
\rho_\delta \delta(z) & 0 & 0 & 0\\
0 & - p_\delta \delta(z) & 0 & 0\\
0 & 0 & - p_\delta \delta(z) & 0\\
0 & 0 & 0 & -p_z \delta(z)
\end{pmatrix}
,
\end{equation}
where $\rho_\delta$ , $ p_\delta$ and $p_z$ are constants.

The $zz$ component of the field equation \eqref{eomfQ3} then gives
\begin{equation}
    \label{eomzz}
    -\frac{f(Q)}{2} + f_{Q} Q =  p_z\delta(z).
\end{equation}
As discussed in the previous subsection,
the nonmetricity scalar $Q$ is constant beyond $z=0$,
as dictated by the specific form of the function $f$.
This leads to two possibilities:
either $Q$ exhibits a removable discontinuity at $z=0$, or $p_z=0$.
The former case indicates that the spacetime may contain a planar defect or brane located at $z=0$.
This is reflected by possible $\delta$ functions in the metric function $u(z)$ or $v(z)$, or their derivatives.
Although the singular matter source described by Eq.\eqref{singularT} already acts as a domain wall,
the presence of a spacetime defect does not necessarily follow.
Nevertheless, as the discontinuity is removable, 
one can still impose continuity of the metric at $z=0$,
which means
\begin{equation}
        v(0^+)=v(0^-),\quad u(0^+)=u(0^-).
\end{equation}
Subsequently, from Eq.\eqref{plusQ0}, we have
\begin{equation}
    \begin{split}
        \left\lvert  \tilde C_{+}\cosh \sqrt{\frac{3Q_0}{8}}A_{+}\right\rvert =&\left\lvert  \tilde C_{-}\cosh \sqrt{\frac{3Q_0}{8}}A_{-}\right\rvert ,\\
        \left\lvert \tilde B_{+} \tilde C_{+}^{\frac{1}{3}}\sinh \sqrt{\frac{3Q_0}{8}}A_{+}\right\rvert =&\left\lvert \tilde B_{-} \tilde C_{-}^{\frac{1}{3}}\sinh \sqrt{\frac{3Q_0}{8}}A_{-}\right\rvert,
    \end{split}
\end{equation}
for $Q_0>0$ and
\begin{equation}
    \begin{split}
        \left\lvert \tilde{C}_{+} \cos \sqrt{\frac{3\left\lvert Q_0\right\rvert }{8}}A_{+} \right\rvert &=\left\lvert \tilde{C}_{-} \cos \sqrt{\frac{3\left\lvert Q_0\right\rvert }{8}}A_{-} \right\rvert ,\\
        \left\lvert \tilde{B}_{+} \tilde{C}^{\frac{1}{3}}_{+}\sin \sqrt{\frac{3\left\lvert Q_0\right\rvert }{8}}A_{+}\right\rvert &=\left\lvert \tilde{B}_{-} \tilde{C}^{\frac{1}{3}}_{-}\sin \sqrt{\frac{3\left\lvert Q_0\right\rvert }{8}}A_{-}\right\rvert,
    \end{split}
\end{equation}
for $Q_0<0$, where $\tilde B_{\pm} = \e^{B_\pm}$, $\tilde C_\pm = \e^{\frac32 C_\pm}$,
and the subscripts $+$ and $-$ of the integration constants $A$, $B$, and $C$
correspond to the regions $z>0$ and $z<0$, respectively.

On the other hand, the $p_z=0$ case appears more natural
in that vanishing perpendicular pressure is consistent with the matter confined at $z=0$.
If this is true and $Q$ equals $Q_0$ uniformly across the $z=0$ plane,
Eq. \eqref{eomfQ3} then reads
\begin{equation}
    \label{eomthinshell}
    \frac{1}{2}g_{\mu\nu}f_{Q_0}Q_0 +f_{Q_0}G_{\mu\nu}=  \mathcal{T}_{\mu\nu},
\end{equation}
where $f_{Q0}$ denotes $f_Q\vert_{Q=Q_0}$.
The $tt$ and $xx$ (or $yy$) components can be recast into
\begin{equation}
    \begin{split}
        v''+\frac{3}{2}v'^2=&\frac{1}{4}Q_0-\frac{\rho_\delta}{2f_{Q_0}}\delta(z),\\
        u''+v''+u'^2+v'^2+u'v'=&\frac{1}{2}Q_0 +\frac{   p_\delta}{f_{Q_0}}\delta(z).
    \end{split}
\end{equation}
Integrating in the vicinity of the $z=0$ plane yields
\begin{equation}
    \begin{split}
        v'(0^{+})-v'(0^{-})&=-\frac{1}{2f_{Q_0}}\rho_\delta,\\
        u'(0^{+})-u'(0^{-})&=\frac{1}{2f_{Q_0}}(\rho_\delta+2 p_\delta),
    \end{split}
\end{equation}
from which it follows that, for $Q_0>0$,
\begin{equation}
    \begin{split}
        \rho_\delta=&-2f_{Q_0}\sqrt{\frac{Q_0}{6}}\left(\tanh\sqrt{\frac{3Q_0}{8}}A_{+}-\tanh\sqrt{\frac{3Q_0}{8}}A_{-}\right),\\ 
         p_\delta=&-\frac{\rho_\delta}{4}+f_{Q_0}\sqrt{\frac{3Q_0}{8}}\left(\coth\sqrt{\frac{3Q_0}{8}}A_{+}-\coth\sqrt{\frac{3Q_0}{8}}A_{-}\right).
    \end{split}
\end{equation}
The dominant energy condition $\rho_\delta>0$ then further requires that either
\begin{equation}
    f_{Q_0}<0,A_{-}<A_{+},
\end{equation}
or
\begin{equation}
    f_{Q_0}>0,A_{-}>A_{+}.
\end{equation}

Similarly, for $Q_0<0$, the energy density and pressure read
\begin{equation}
    \begin{split}
        \rho_\delta&=2f_{Q_0}\sqrt{\frac{|Q_0|}{6}}\left(\tan\sqrt{\frac{3\left\lvert Q_0\right\rvert }{8}}A_{+}-\tan\sqrt{\frac{3\left\lvert Q_0\right\rvert }{8}}A_{-}\right) ,\\
         p_\delta&=-\frac{\rho_\delta}{4}+f_{Q_0}\sqrt{\frac{3\left\lvert Q_0\right\rvert }{8}}\left(\cot\sqrt{\frac{3\left\lvert Q_0\right\rvert }{8}}A_{+}-\cot\sqrt{\frac{3\left\lvert Q_0\right\rvert }{8}}A_{-}\right).
    \end{split}
\end{equation}
And the condition $\rho_\delta>0$ yields
\begin{equation}
    f_{Q_0}<0,\quad\tan\sqrt{\frac{3\left\lvert Q_0\right\rvert }{8}}A_{+}<\tan\sqrt{\frac{3\left\lvert Q_0\right\rvert }{8}}A_{-}
\end{equation}
or
\begin{equation}
    f_{Q_0}>0,\quad\tan\sqrt{\frac{3\left\lvert Q_0\right\rvert }{8}}A_{-}<\tan\sqrt{\frac{3\left\lvert Q_0\right\rvert }{8}}A_{+}.
\end{equation}

\section{Matter slabs of finite thickness}
\label{slab}
In this section, we explore static plane-symmetric matter slabs of finite thickness, which
may serve as non-singular sources for the external vacuum geometry derived earlier.
For simplicity, we focus on an isotropic fluid with a constant positive density $\rho_0$.
The energy-momentum tensor for such a fluid takes the form
\begin{equation}
\mathcal{T}^{\mu}_{\;\;\;\;\nu}=
\begin{pmatrix}
\rho_0  & 0 & 0 & 0\\
0 & -p(z) & 0 & 0\\
0 & 0 & -p(z) & 0\\
0 & 0 & 0 & -p(z)
\end{pmatrix}
,
\end{equation}
where the isotropic pressure $p(z)$ depends only on the $z$ coordinate.
The conservation of $\mathcal T_{\mu\nu}$ immediately yields
\begin{equation}
    \label{nablaT}
    p'(z)+u'(z)\left(p(z)+\rho_0\right)=0.
\end{equation}
Note that among the four equations comprising this relation and
all three components of Eq.\eqref{Peom},
only three are independent due to the contracted Bianchi identity.
We therefore adopt the $tt$ and $zz$ components of Eq.\eqref{Peom} together with Eq.\eqref{nablaT} as the independent set.
Substituting Eq.\eqref{nablaT} into the other two equations,
one can eliminate all derivatives of $u(z)$,
producing a closed system of two second-order differential equations for $p(z)$ and $v(z)$,
provided that the function $f(Q)$ is specified.
The resulting system formally requires four boundary conditions.
However, since $v(z)$ does not explicitly appear in the equations and, thus,
the system is invariant under a constant shift in $v(z)$ itself,
only three independent boundary conditions are required to determine a unique physical solution within the finite-thickness matter slab.

As an illustrative example, we examine a simple model \cite{Lin:2021uqa}
\begin{equation}
    f(Q)=Q+\alpha Q^2,
\end{equation}
with $\alpha$ being the model parameter of dimension length squared.
In the external vacuum, as discussed in the previous section,
this yields an algebraic solution and fixes the value of $Q=Q_0=-\frac1{3\alpha}$.
For the slab to form a self-gravitating system with naturally occurring boundaries,
the pressure perpendicular to the surfaces must vanish, i.e., $p(z_{\pm})=0$,
where $z_{\pm}$ denote the $z$ coordinates of the two surfaces of the slab.
As for the boundary conditions for $v(z)$,
the junction conditions at the surface $z=z_\pm$ require that
the interior metric match the external vacuum solution.
Specifically, from Eq.\eqref{PQ}, 
the condition $Q(z_\pm)=Q_0$ implies that $u'(z_\pm)$ and $v'(z_\pm)$ are mutually dependent.
For the pressure to build up to positive values inward from the zero-pressure surfaces, 
the solution must satisfy $u'(z_-) < 0$, 
as this is a necessary condition derived from the relation between $u(z)$ and $p(z)$.
We therefore choose $u_0'=u'(z_-)$ as a free parameter.
Collecting the pieces, we now have the complete set of three boundary needed,
two on $p(z_\pm)$ from the requirement of natural surfaces
and the third on $v'(z_-)$ from the free parameter $u_0'$ via Eq.\eqref{PQ},
to determine the differential system derived previously.
One can relate $u_0'$ to the integration constants of the vacuum region through
\begin{equation}
    \label{numerParameter}
    u_0'=u'(z_-)=\sqrt{\frac1{8|\alpha|}}\left[\coth\sqrt{\frac1{8|\alpha|}}(z_-+A_-)-\frac13\tanh\sqrt{\frac1{8|\alpha|}(z_-+A_-)}\right].
\end{equation}

Numerical procedure confirms that, with appropriate choices of parameters,
the system given by Eqs.\eqref{eomfQ3} and \eqref{nablaT} is able to converge to physically consistent 
solutions describing slabs of finite thickness with two well-defined surfaces.

\begin{figure}[htpb]
    \centering
    \includegraphics[width=0.8\linewidth]{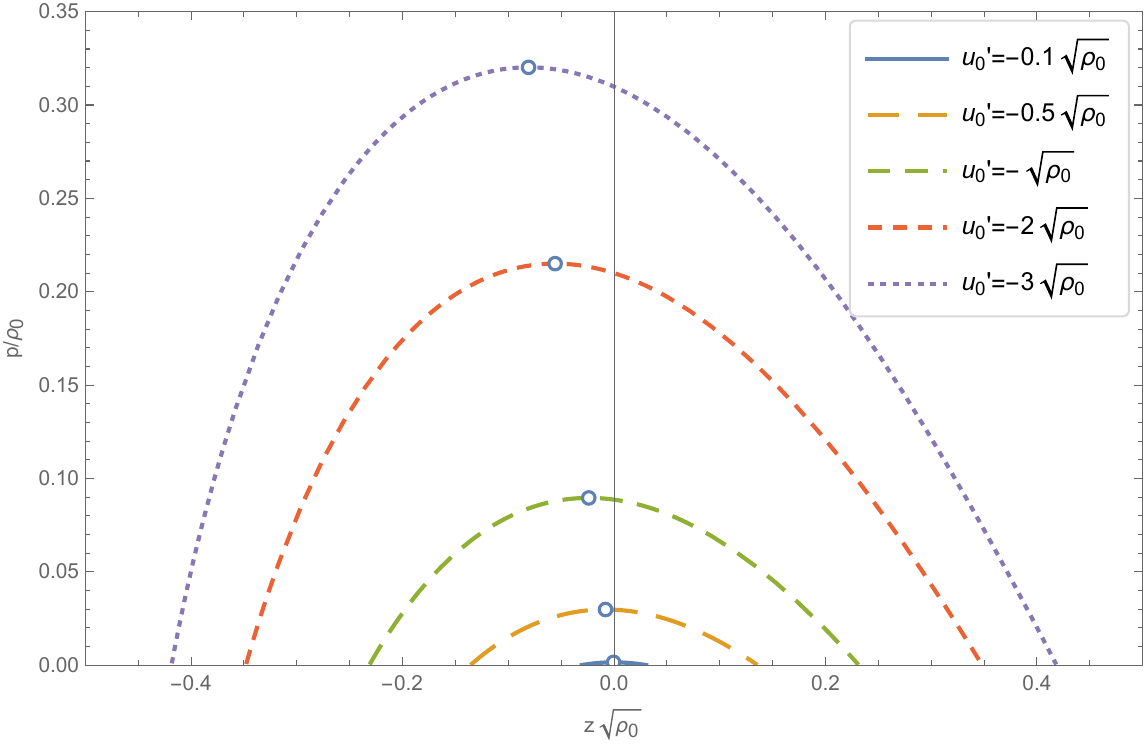}
    \caption{The profiles of the isotropic pressure $p$
    for $\alpha=-0.1\rho_0^{-1}$ and various $u'(z_-)$.
    The peaks are located at $(-0.004,0.0015)$, $(-0.008,0.028)$, $(-0.024,0.089)$, $(-0.056,0.215)$, $(-0.081,0.320)$,
    for $u_0'/\sqrt{\rho_0}=-0.1$, $-0.5$, $-1$, $-2$, $-3$, respectively,
    where the geometric center $z=(z_-+z_+)/2$ has been moved to $z=0$.}
    \label{fig:a01}
\end{figure}
Figure \ref{fig:a01} illustrates the pressure profiles for $\alpha=-0.1\rho_0^{-1}$ and various $u'(z_-)$.
Since $-u'(z_-)$ characterizes the gradient of pressure increase from the surface,
the pressure is observed to be higher for larger magnitudes of $|u_0'|$.
It is also evident that the profiles of pressure are not symmetric about $z=\left(z_++z_-\right)/2$.
This asymmetry arises because the condition $Q(z_+) = Q_0= Q(z_-)$ does not constrain $u'(z_+)$ solely, 
as both $u'(z)$ and $v'(z)$ contribute to $Q(z)$. 
Consequently, at the point $z=z_+$ where $u(z_+)=u(z_-)$ so that $p(z_+)=0=p(z_-)$,
the value $-u'(z_+)$ generally differs from the chosen parameter $u_0'=u'(z_-)$,
which accounts for the asymmetric pressure behavior.
Despite this asymmetry, the system admits solutions that are mirror-reflected across the $z=(z_-+z_+)/2$ plane.
Such flipped solutions can be obtained by initiating the integration at $z = z_+$ and solving toward $z = z_-$.

\begin{figure}[htpb]
    \centering
    \includegraphics[width=0.8\linewidth]{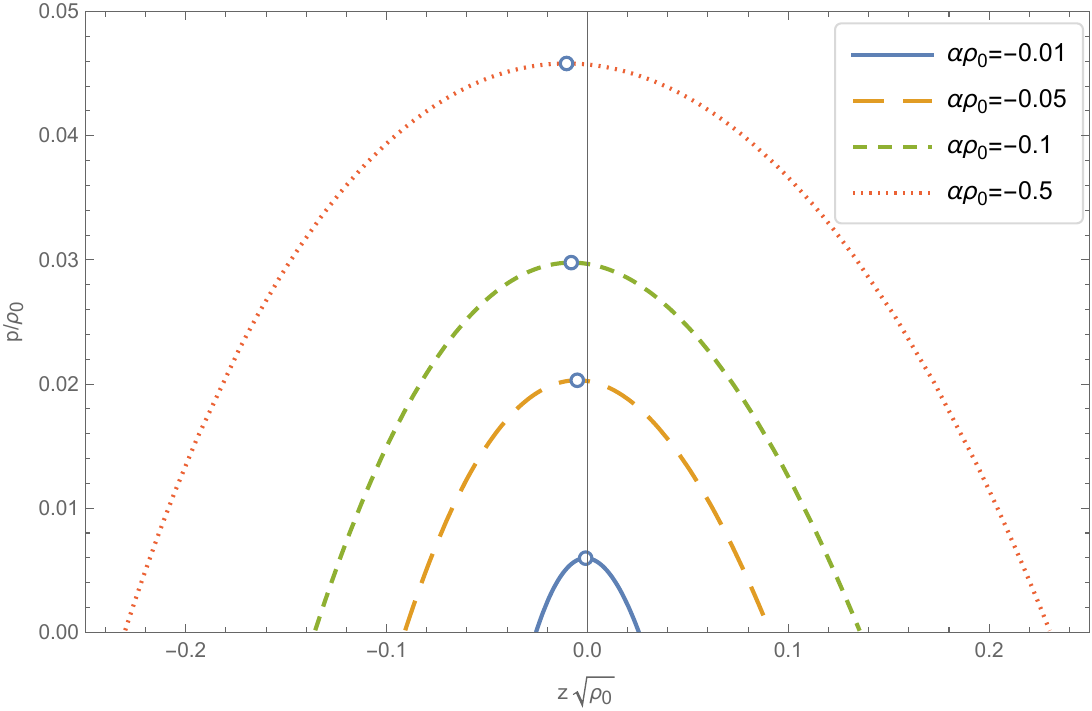}
    \caption{The profiles of the isotropic pressure $p$
    for $u'(z_-)=-0.1\sqrt{\rho_0}$ and various $\alpha$.
    The peaks are located at $(-0.0008,0.006)$, $(-0.005,0.020)$, $(-0.008,0.029)$, $(-0.010,0.046)$,
    for $\alpha\rho_0=-0.01$, $-0.05$, $-0.1$, $-0.5$, respectively
    where the geometric center $z=(z_-+z_+)/2$ has been moved to $z=0$.}
    \label{fig:u01}
\end{figure}
In Fig. \ref{fig:u01}, we present the pressure profiles for $u'(z_-)=-0.1\sqrt{\rho_0}$ and various $\alpha$.
The thickness of the slab and the pressure within generally increase with $|\alpha|$.
However, the GR case with $\alpha=0$ does not follow such a pattern.
In fact, the procedure $\alpha\rightarrow0$ is not continuous here.
This discontinuity finds a parallel in the vacuum case, 
where, as discussed in Sect. \ref{vac}, the standard Taub solution cannot be recovered by simply setting $Q_0 = 0$.
A similar reasoning applies here:
the cases $\alpha=0$ and $\alpha<0$ belong to distinct branches of the equations for $u'(z)$ and $v'(z)$,
thereby precluding a smooth transition between them.

However, the numerical procedure fails to converge for $\alpha>0$.
This can be understood as follows. 
For a slab with two natural surfaces at $z_\pm$ where $p(z_\pm)=0$, 
the pressure profile $p(z)$ must possess at least one extremum at an interior point $z_m \in (z_-, z_+)$,
which, from Eq. \eqref{nablaT}, means $u'(z_m)=0$, and consequently, $Q(z_m) = 2v'(z_m)^2 \ge 0$. 
On the other hand, the $zz$ component of the field equation can be recast into
\begin{equation}
    Q(z)=-\frac1{6\alpha}\left(1\pm\sqrt{1+24\alpha p(z)}\right),
\end{equation}
which reveals that $Q(z)$ has two disconnected regimes for $p(z)\ge 0$ and $\alpha>0$.
Crucially, the junction conditions at the surfaces require $Q(z_\pm) = Q_0 = -\frac{1}{3\alpha} < 0$ for $\alpha>0$. 
This forces $Q(z)$ to remain in the negative regime.
Thus, the necessity $Q(z_m)\ge0$ cannot be satisfied,
which precludes the existence of a consistent solution with two natural surfaces for $\alpha>0$.

\section{Conclusion and discussions}
\label{conclusion}
In this paper, we investigate the static plane symmetric configurations within the framework of $f(Q)$ gravity.
We derive vacuum solutions for general $f(Q)$ models 
and study two kinds of matter distribution,
i.e., singular thin shell and slab of finite thickness,
which may serve as the source of such vacuum geometries.
In vacuum case, one of the field equations reduces to an algebraic equation of $Q$,
provided that the functional form of $f(Q)$ is specified and regular.
This results in a fixed value $Q=Q_0$ throughout the vacuum region once $f(Q)$ is given. 
It follows that the field equations are greatly simplified,
allowing us to obtain a series of vacuum solutions corresponding to the Taub-(anti-)de Sitter solutions.
If a thin, infinite material plane situates at $z=0$ acting as the source of these solutions,
the vacuum regions $z>0$ and $z<0$ may be described by some geometry (either Taub-de Sitter or Taub-anti de Sitter)
with different integration constants that are related to the energy density and pressure of the matter plane.
Different choices of $f(Q)$ only affect the constant $Q_0$ effectively acting as the cosmological constant.
This behavior arises naturally in $f(Q)$ gravity, 
in contrast to $f(R)$ or $f(T)$ theories where similar conditions are often imposed by hand.
It is most likely a consequence of the high symmetry inherent in static planar configuration
combined with the vanishing connection ans\"atze.
Notably, unlike the case in the spherical symmetry in $f(Q)$ gravity \cite{Lin:2021uqa}
where the nonmetricity scalar $Q$ always vanishes,
the constant $Q=Q_0$ may be different for different forms of $f(Q)$ in plane symmetry.
One may expect that this constancy of $Q$ would not persist in scenarios with lesser symmetry
or with an alternative choice of connection.

We further consider the case where vacuum regions are sourced by a matter slab of finite thickness. 
In this context, the internal structure of the slab does not degenerate across different $f(Q)$ models.
For solving the interior field equations, well-defined vacuums with static plane symmetry offer reliable junction conditions, 
which serve as appropriate boundary conditions. 
The requirement of natural surfaces with zero perpendicular pressures, 
together with the positivity of pressures, further constrains the parameter freedom.
Yet there are still some free parameters that may affect the pressure profiles.
We demonstrate these effects of $f(Q)$ forms and boundary parameters by considering
a quadratic model of $f(Q)=Q+\alpha Q^2$ and an isotropic matter slab with constant energy density.
Through numerical procedures,
we show that the peaks of material pressures within the slab generally are not located at the geometric center.
This pressure asymmetry then admits metric configurations with different integration constants in the two disconnected vacuum regions
even though both share the same universal value $Q = Q_0$.
Moreover, we find that a negative $\alpha$ with larger $|\alpha|$ results in generally larger pressure and thickness.
This can be interpreted as a weakening of gravity or 
as the presence of an effective geometric fluid 
that provides additional pressure to counteract gravitational collapse, 
thereby supporting more matter in a thicker configuration.
In contrast, models with positive $\alpha$ fail to form natural surfaces. 
This stems from an inherent contradiction in the required behaviors of the function $Q(z)$,
which should be non-negative for pressure to have a peak within the slab,
while it also needs to remain in the negative regime for natural surface to have $p(z_\pm)=0$.
This conclusion can be generalized to some other models, e.g., $f(Q)=Q+\alpha Q^4$ or $f(Q)=Q\e^{\alpha Q}$,
where $f(Q)>Q$ and $f'(Q)>1$ for $\alpha>0$, leading to $Q_0<0$ at the surface.

The universal constancy of the nonmetricity scalar $Q$ in the vacuum is crucial to the current study.
As an algebraic constraint arising from the functional form of $f(Q)$,
it restricts the vacuum geometries to be the same kind (either Taub-de Sitter or Taub-anti de Sitter), 
even the vacuum regions may not be connected.
This constraint also necessitates consistent boundary conditions for any interior matter source, 
as it must join regions of the same kind of vacuum geometry.
Nevertheless, some interesting cases may have been omitted in our analysis.
If Eq.\eqref{PAlgebeq} admits multiple algebraic roots,
particularly ones with opposite signs,
different kinds of vacuum geometries could in principle coexist.
Beyond this algebraic ambiguity, 
the choice of the affine connection itself presents further degrees of freedom. 
Specifically, the symmetry equations \eqref{lieD} and torsion-free condition
leave 12 components of the connection undetermined, 
which should be subjected to the vanishing curvature condition.
Solving the full system may still leave residual freedom, potentially allowing for a non-constant $Q$. 
These aspects merit further investigation, 
as they may illuminate deeper effects of the $f(Q)$ modification to gravity.

\section*{Acknowledgement}
\label{ackn}
This work is supported by the National Science Foundation of China under Grant No. 12575060.

\bibliography{ref}

\begin{thebibliography}{63}%
\makeatletter
\providecommand \@ifxundefined [1]{%
 \@ifx{#1\undefined}
}%
\providecommand \@ifnum [1]{%
 \ifnum #1\expandafter \@firstoftwo
 \else \expandafter \@secondoftwo
 \fi
}%
\providecommand \@ifx [1]{%
 \ifx #1\expandafter \@firstoftwo
 \else \expandafter \@secondoftwo
 \fi
}%
\providecommand \natexlab [1]{#1}%
\providecommand \enquote  [1]{``#1''}%
\providecommand \bibnamefont  [1]{#1}%
\providecommand \bibfnamefont [1]{#1}%
\providecommand \citenamefont [1]{#1}%
\providecommand \href@noop [0]{\@secondoftwo}%
\providecommand \href [0]{\begingroup \@sanitize@url \@href}%
\providecommand \@href[1]{\@@startlink{#1}\@@href}%
\providecommand \@@href[1]{\endgroup#1\@@endlink}%
\providecommand \@sanitize@url [0]{\catcode `\\12\catcode `\$12\catcode `\&12\catcode `\#12\catcode `\^12\catcode `\_12\catcode `\%12\relax}%
\providecommand \@@startlink[1]{}%
\providecommand \@@endlink[0]{}%
\providecommand \url  [0]{\begingroup\@sanitize@url \@url }%
\providecommand \@url [1]{\endgroup\@href {#1}{\urlprefix }}%
\providecommand \urlprefix  [0]{URL }%
\providecommand \Eprint [0]{\href }%
\providecommand \doibase [0]{https://doi.org/}%
\providecommand \selectlanguage [0]{\@gobble}%
\providecommand \bibinfo  [0]{\@secondoftwo}%
\providecommand \bibfield  [0]{\@secondoftwo}%
\providecommand \translation [1]{[#1]}%
\providecommand \BibitemOpen [0]{}%
\providecommand \bibitemStop [0]{}%
\providecommand \bibitemNoStop [0]{.\EOS\space}%
\providecommand \EOS [0]{\spacefactor3000\relax}%
\providecommand \BibitemShut  [1]{\csname bibitem#1\endcsname}%
\let\auto@bib@innerbib\@empty
\bibitem [{\citenamefont {Beltr{\'a}n~Jim{\'e}nez}\ \emph {et~al.}(2019)\citenamefont {Beltr{\'a}n~Jim{\'e}nez}, \citenamefont {Heisenberg},\ and\ \citenamefont {Koivisto}}]{BeltranJimenez:2019esp}%
  \BibitemOpen
  \bibfield  {author} {\bibinfo {author} {\bibfnamefont {J.}~\bibnamefont {Beltr{\'a}n~Jim{\'e}nez}}, \bibinfo {author} {\bibfnamefont {L.}~\bibnamefont {Heisenberg}},\ and\ \bibinfo {author} {\bibfnamefont {T.~S.}\ \bibnamefont {Koivisto}},\ }\bibfield  {title} {\bibinfo {title} {{The Geometrical Trinity of Gravity}},\ }\href {https://doi.org/10.3390/universe5070173} {\bibfield  {journal} {\bibinfo  {journal} {Universe}\ }\textbf {\bibinfo {volume} {5}},\ \bibinfo {pages} {173} (\bibinfo {year} {2019})},\ \Eprint {https://arxiv.org/abs/1903.06830} {arXiv:1903.06830 [hep-th]} \BibitemShut {NoStop}%
\bibitem [{\citenamefont {Harada}(2020)}]{PhysRevD.101.024053}%
  \BibitemOpen
  \bibfield  {author} {\bibinfo {author} {\bibfnamefont {J.}~\bibnamefont {Harada}},\ }\bibfield  {title} {\bibinfo {title} {Connection independent formulation of general relativity},\ }\href {https://doi.org/10.1103/PhysRevD.101.024053} {\bibfield  {journal} {\bibinfo  {journal} {Phys. Rev. D}\ }\textbf {\bibinfo {volume} {101}},\ \bibinfo {pages} {024053} (\bibinfo {year} {2020})}\BibitemShut {NoStop}%
\bibitem [{\citenamefont {Maluf}(2013)}]{Maluf2013}%
  \BibitemOpen
  \bibfield  {author} {\bibinfo {author} {\bibfnamefont {J.~W.}\ \bibnamefont {Maluf}},\ }\bibfield  {title} {\bibinfo {title} {The teleparallel equivalent of general relativity},\ }\href {https://doi.org/https://doi.org/10.1002/andp.201200272} {\bibfield  {journal} {\bibinfo  {journal} {Annalen der Physik}\ }\textbf {\bibinfo {volume} {525}},\ \bibinfo {pages} {339} (\bibinfo {year} {2013})},\ \Eprint {https://arxiv.org/abs/https://onlinelibrary.wiley.com/doi/pdf/10.1002/andp.201200272} {https://onlinelibrary.wiley.com/doi/pdf/10.1002/andp.201200272} \BibitemShut {NoStop}%
\bibitem [{\citenamefont {Aldrovandi}\ and\ \citenamefont {Pereira}(2013)}]{Aldrovandi:2013wha}%
  \BibitemOpen
  \bibfield  {author} {\bibinfo {author} {\bibfnamefont {R.}~\bibnamefont {Aldrovandi}}\ and\ \bibinfo {author} {\bibfnamefont {J.~G.}\ \bibnamefont {Pereira}},\ }\href {https://doi.org/10.1007/978-94-007-5143-9} {\emph {\bibinfo {title} {{Teleparallel Gravity}: {An Introduction}}}},\ Vol.\ \bibinfo {volume} {173}\ (\bibinfo  {publisher} {Springer},\ \bibinfo {year} {2013})\BibitemShut {NoStop}%
\bibitem [{\citenamefont {Nester}\ and\ \citenamefont {Yo}(1999)}]{Nester:1998mp}%
  \BibitemOpen
  \bibfield  {author} {\bibinfo {author} {\bibfnamefont {J.~M.}\ \bibnamefont {Nester}}\ and\ \bibinfo {author} {\bibfnamefont {H.-J.}\ \bibnamefont {Yo}},\ }\bibfield  {title} {\bibinfo {title} {{Symmetric teleparallel general relativity}},\ }\href@noop {} {\bibfield  {journal} {\bibinfo  {journal} {Chin. J. Phys.}\ }\textbf {\bibinfo {volume} {37}},\ \bibinfo {pages} {113} (\bibinfo {year} {1999})},\ \Eprint {https://arxiv.org/abs/gr-qc/9809049} {arXiv:gr-qc/9809049} \BibitemShut {NoStop}%
\bibitem [{\citenamefont {Adak}\ \emph {et~al.}(2006)\citenamefont {Adak}, \citenamefont {Kalay},\ and\ \citenamefont {Sert}}]{Adak:2005cd}%
  \BibitemOpen
  \bibfield  {author} {\bibinfo {author} {\bibfnamefont {M.}~\bibnamefont {Adak}}, \bibinfo {author} {\bibfnamefont {M.}~\bibnamefont {Kalay}},\ and\ \bibinfo {author} {\bibfnamefont {O.}~\bibnamefont {Sert}},\ }\bibfield  {title} {\bibinfo {title} {{Lagrange formulation of the symmetric teleparallel gravity}},\ }\href {https://doi.org/10.1142/S0218271806008474} {\bibfield  {journal} {\bibinfo  {journal} {Int. J. Mod. Phys. D}\ }\textbf {\bibinfo {volume} {15}},\ \bibinfo {pages} {619} (\bibinfo {year} {2006})},\ \Eprint {https://arxiv.org/abs/gr-qc/0505025} {arXiv:gr-qc/0505025} \BibitemShut {NoStop}%
\bibitem [{\citenamefont {Mol}(2017)}]{Mol:2014ooa}%
  \BibitemOpen
  \bibfield  {author} {\bibinfo {author} {\bibfnamefont {I.}~\bibnamefont {Mol}},\ }\bibfield  {title} {\bibinfo {title} {{The Non-Metricity Formulation of General Relativity}},\ }\href {https://doi.org/10.1007/s00006-016-0749-8} {\bibfield  {journal} {\bibinfo  {journal} {Adv. Appl. Clifford Algebras}\ }\textbf {\bibinfo {volume} {27}},\ \bibinfo {pages} {2607} (\bibinfo {year} {2017})},\ \Eprint {https://arxiv.org/abs/1406.0737} {arXiv:1406.0737 [gr-qc]} \BibitemShut {NoStop}%
\bibitem [{\citenamefont {Jim\'enez}\ \emph {et~al.}(2018)\citenamefont {Jim\'enez}, \citenamefont {Heisenberg},\ and\ \citenamefont {Koivisto}}]{PhysRevD.98.044048}%
  \BibitemOpen
  \bibfield  {author} {\bibinfo {author} {\bibfnamefont {J.~B.}\ \bibnamefont {Jim\'enez}}, \bibinfo {author} {\bibfnamefont {L.}~\bibnamefont {Heisenberg}},\ and\ \bibinfo {author} {\bibfnamefont {T.}~\bibnamefont {Koivisto}},\ }\bibfield  {title} {\bibinfo {title} {Coincident general relativity},\ }\href {https://doi.org/10.1103/PhysRevD.98.044048} {\bibfield  {journal} {\bibinfo  {journal} {Phys. Rev. D}\ }\textbf {\bibinfo {volume} {98}},\ \bibinfo {pages} {044048} (\bibinfo {year} {2018})}\BibitemShut {NoStop}%
\bibitem [{\citenamefont {Sotiriou}\ and\ \citenamefont {Faraoni}(2010)}]{Sotiriou:2008rp}%
  \BibitemOpen
  \bibfield  {author} {\bibinfo {author} {\bibfnamefont {T.~P.}\ \bibnamefont {Sotiriou}}\ and\ \bibinfo {author} {\bibfnamefont {V.}~\bibnamefont {Faraoni}},\ }\bibfield  {title} {\bibinfo {title} {{f(R) Theories Of Gravity}},\ }\href {https://doi.org/10.1103/RevModPhys.82.451} {\bibfield  {journal} {\bibinfo  {journal} {Rev. Mod. Phys.}\ }\textbf {\bibinfo {volume} {82}},\ \bibinfo {pages} {451} (\bibinfo {year} {2010})},\ \Eprint {https://arxiv.org/abs/0805.1726} {arXiv:0805.1726 [gr-qc]} \BibitemShut {NoStop}%
\bibitem [{\citenamefont {De~Felice}\ and\ \citenamefont {Tsujikawa}(2010)}]{DeFelice:2010aj}%
  \BibitemOpen
  \bibfield  {author} {\bibinfo {author} {\bibfnamefont {A.}~\bibnamefont {De~Felice}}\ and\ \bibinfo {author} {\bibfnamefont {S.}~\bibnamefont {Tsujikawa}},\ }\bibfield  {title} {\bibinfo {title} {{f(R) theories}},\ }\href {https://doi.org/10.12942/lrr-2010-3} {\bibfield  {journal} {\bibinfo  {journal} {Living Rev. Rel.}\ }\textbf {\bibinfo {volume} {13}},\ \bibinfo {pages} {3} (\bibinfo {year} {2010})},\ \Eprint {https://arxiv.org/abs/1002.4928} {arXiv:1002.4928 [gr-qc]} \BibitemShut {NoStop}%
\bibitem [{\citenamefont {Nojiri}\ and\ \citenamefont {Odintsov}(2011)}]{Nojiri:2010wj}%
  \BibitemOpen
  \bibfield  {author} {\bibinfo {author} {\bibfnamefont {S.}~\bibnamefont {Nojiri}}\ and\ \bibinfo {author} {\bibfnamefont {S.~D.}\ \bibnamefont {Odintsov}},\ }\bibfield  {title} {\bibinfo {title} {{Unified cosmic history in modified gravity: from F(R) theory to Lorentz non-invariant models}},\ }\href {https://doi.org/10.1016/j.physrep.2011.04.001} {\bibfield  {journal} {\bibinfo  {journal} {Phys. Rept.}\ }\textbf {\bibinfo {volume} {505}},\ \bibinfo {pages} {59} (\bibinfo {year} {2011})},\ \Eprint {https://arxiv.org/abs/1011.0544} {arXiv:1011.0544 [gr-qc]} \BibitemShut {NoStop}%
\bibitem [{\citenamefont {Capozziello}\ and\ \citenamefont {De~Laurentis}(2011)}]{Capozziello:2011et}%
  \BibitemOpen
  \bibfield  {author} {\bibinfo {author} {\bibfnamefont {S.}~\bibnamefont {Capozziello}}\ and\ \bibinfo {author} {\bibfnamefont {M.}~\bibnamefont {De~Laurentis}},\ }\bibfield  {title} {\bibinfo {title} {{Extended Theories of Gravity}},\ }\href {https://doi.org/10.1016/j.physrep.2011.09.003} {\bibfield  {journal} {\bibinfo  {journal} {Phys. Rept.}\ }\textbf {\bibinfo {volume} {509}},\ \bibinfo {pages} {167} (\bibinfo {year} {2011})},\ \Eprint {https://arxiv.org/abs/1108.6266} {arXiv:1108.6266 [gr-qc]} \BibitemShut {NoStop}%
\bibitem [{\citenamefont {Cai}\ \emph {et~al.}(2016)\citenamefont {Cai}, \citenamefont {Capozziello}, \citenamefont {De~Laurentis},\ and\ \citenamefont {Saridakis}}]{Cai:2015emx}%
  \BibitemOpen
  \bibfield  {author} {\bibinfo {author} {\bibfnamefont {Y.-F.}\ \bibnamefont {Cai}}, \bibinfo {author} {\bibfnamefont {S.}~\bibnamefont {Capozziello}}, \bibinfo {author} {\bibfnamefont {M.}~\bibnamefont {De~Laurentis}},\ and\ \bibinfo {author} {\bibfnamefont {E.~N.}\ \bibnamefont {Saridakis}},\ }\bibfield  {title} {\bibinfo {title} {{f(T) teleparallel gravity and cosmology}},\ }\href {https://doi.org/10.1088/0034-4885/79/10/106901} {\bibfield  {journal} {\bibinfo  {journal} {Rept. Prog. Phys.}\ }\textbf {\bibinfo {volume} {79}},\ \bibinfo {pages} {106901} (\bibinfo {year} {2016})},\ \Eprint {https://arxiv.org/abs/1511.07586} {arXiv:1511.07586 [gr-qc]} \BibitemShut {NoStop}%
\bibitem [{\citenamefont {Nojiri}\ \emph {et~al.}(2017)\citenamefont {Nojiri}, \citenamefont {Odintsov},\ and\ \citenamefont {Oikonomou}}]{Nojiri:2017ncd}%
  \BibitemOpen
  \bibfield  {author} {\bibinfo {author} {\bibfnamefont {S.}~\bibnamefont {Nojiri}}, \bibinfo {author} {\bibfnamefont {S.~D.}\ \bibnamefont {Odintsov}},\ and\ \bibinfo {author} {\bibfnamefont {V.~K.}\ \bibnamefont {Oikonomou}},\ }\bibfield  {title} {\bibinfo {title} {{Modified Gravity Theories on a Nutshell: Inflation, Bounce and Late-time Evolution}},\ }\href {https://doi.org/10.1016/j.physrep.2017.06.001} {\bibfield  {journal} {\bibinfo  {journal} {Phys. Rept.}\ }\textbf {\bibinfo {volume} {692}},\ \bibinfo {pages} {1} (\bibinfo {year} {2017})},\ \Eprint {https://arxiv.org/abs/1705.11098} {arXiv:1705.11098 [gr-qc]} \BibitemShut {NoStop}%
\bibitem [{\citenamefont {Lin}\ \emph {et~al.}(2022)\citenamefont {Lin}, \citenamefont {Chen},\ and\ \citenamefont {Zhai}}]{Lin:2021ijx}%
  \BibitemOpen
  \bibfield  {author} {\bibinfo {author} {\bibfnamefont {R.-H.}\ \bibnamefont {Lin}}, \bibinfo {author} {\bibfnamefont {X.-N.}\ \bibnamefont {Chen}},\ and\ \bibinfo {author} {\bibfnamefont {X.-H.}\ \bibnamefont {Zhai}},\ }\bibfield  {title} {\bibinfo {title} {{Realistic neutron star models in f(T) gravity}},\ }\href {https://doi.org/10.1140/epjc/s10052-022-10268-2} {\bibfield  {journal} {\bibinfo  {journal} {Eur. Phys. J. C}\ }\textbf {\bibinfo {volume} {82}},\ \bibinfo {pages} {308} (\bibinfo {year} {2022})},\ \Eprint {https://arxiv.org/abs/2109.00191} {arXiv:2109.00191 [gr-qc]} \BibitemShut {NoStop}%
\bibitem [{\citenamefont {Yang}\ \emph {et~al.}(2022)\citenamefont {Yang}, \citenamefont {Lin},\ and\ \citenamefont {Zhai}}]{Yang:2022efz}%
  \BibitemOpen
  \bibfield  {author} {\bibinfo {author} {\bibfnamefont {J.}~\bibnamefont {Yang}}, \bibinfo {author} {\bibfnamefont {R.-H.}\ \bibnamefont {Lin}},\ and\ \bibinfo {author} {\bibfnamefont {X.-H.}\ \bibnamefont {Zhai}},\ }\bibfield  {title} {\bibinfo {title} {{Viscous cosmology in f(T) gravity}},\ }\href {https://doi.org/10.1140/epjc/s10052-022-11008-2} {\bibfield  {journal} {\bibinfo  {journal} {Eur. Phys. J. C}\ }\textbf {\bibinfo {volume} {82}},\ \bibinfo {pages} {1039} (\bibinfo {year} {2022})},\ \Eprint {https://arxiv.org/abs/2208.09991} {arXiv:2208.09991 [gr-qc]} \BibitemShut {NoStop}%
\bibitem [{\citenamefont {Beltr{\'a}n~Jim{\'e}nez}\ \emph {et~al.}(2020)\citenamefont {Beltr{\'a}n~Jim{\'e}nez}, \citenamefont {Heisenberg}, \citenamefont {Koivisto},\ and\ \citenamefont {Pekar}}]{BeltranJimenez:2019tme}%
  \BibitemOpen
  \bibfield  {author} {\bibinfo {author} {\bibfnamefont {J.}~\bibnamefont {Beltr{\'a}n~Jim{\'e}nez}}, \bibinfo {author} {\bibfnamefont {L.}~\bibnamefont {Heisenberg}}, \bibinfo {author} {\bibfnamefont {T.~S.}\ \bibnamefont {Koivisto}},\ and\ \bibinfo {author} {\bibfnamefont {S.}~\bibnamefont {Pekar}},\ }\bibfield  {title} {\bibinfo {title} {{Cosmology in $f(Q)$ geometry}},\ }\href {https://doi.org/10.1103/PhysRevD.101.103507} {\bibfield  {journal} {\bibinfo  {journal} {Phys. Rev. D}\ }\textbf {\bibinfo {volume} {101}},\ \bibinfo {pages} {103507} (\bibinfo {year} {2020})},\ \Eprint {https://arxiv.org/abs/1906.10027} {arXiv:1906.10027 [gr-qc]} \BibitemShut {NoStop}%
\bibitem [{\citenamefont {Lazkoz}\ \emph {et~al.}(2019)\citenamefont {Lazkoz}, \citenamefont {Lobo}, \citenamefont {Ortiz-Ba{\~n}os},\ and\ \citenamefont {Salzano}}]{Lazkoz:2019sjl}%
  \BibitemOpen
  \bibfield  {author} {\bibinfo {author} {\bibfnamefont {R.}~\bibnamefont {Lazkoz}}, \bibinfo {author} {\bibfnamefont {F.~S.~N.}\ \bibnamefont {Lobo}}, \bibinfo {author} {\bibfnamefont {M.}~\bibnamefont {Ortiz-Ba{\~n}os}},\ and\ \bibinfo {author} {\bibfnamefont {V.}~\bibnamefont {Salzano}},\ }\bibfield  {title} {\bibinfo {title} {{Observational constraints of $f(Q)$ gravity}},\ }\href {https://doi.org/10.1103/PhysRevD.100.104027} {\bibfield  {journal} {\bibinfo  {journal} {Phys. Rev. D}\ }\textbf {\bibinfo {volume} {100}},\ \bibinfo {pages} {104027} (\bibinfo {year} {2019})},\ \Eprint {https://arxiv.org/abs/1907.13219} {arXiv:1907.13219 [gr-qc]} \BibitemShut {NoStop}%
\bibitem [{\citenamefont {Lu}\ \emph {et~al.}(2019)\citenamefont {Lu}, \citenamefont {Zhao},\ and\ \citenamefont {Chee}}]{Lu:2019hra}%
  \BibitemOpen
  \bibfield  {author} {\bibinfo {author} {\bibfnamefont {J.}~\bibnamefont {Lu}}, \bibinfo {author} {\bibfnamefont {X.}~\bibnamefont {Zhao}},\ and\ \bibinfo {author} {\bibfnamefont {G.}~\bibnamefont {Chee}},\ }\bibfield  {title} {\bibinfo {title} {{Cosmology in symmetric teleparallel gravity and its dynamical system}},\ }\href {https://doi.org/10.1140/epjc/s10052-019-7038-3} {\bibfield  {journal} {\bibinfo  {journal} {Eur. Phys. J. C}\ }\textbf {\bibinfo {volume} {79}},\ \bibinfo {pages} {530} (\bibinfo {year} {2019})},\ \Eprint {https://arxiv.org/abs/1906.08920} {arXiv:1906.08920 [gr-qc]} \BibitemShut {NoStop}%
\bibitem [{\citenamefont {Barros}\ \emph {et~al.}(2020)\citenamefont {Barros}, \citenamefont {Barreiro}, \citenamefont {Koivisto},\ and\ \citenamefont {Nunes}}]{Barros:2020bgg}%
  \BibitemOpen
  \bibfield  {author} {\bibinfo {author} {\bibfnamefont {B.~J.}\ \bibnamefont {Barros}}, \bibinfo {author} {\bibfnamefont {T.}~\bibnamefont {Barreiro}}, \bibinfo {author} {\bibfnamefont {T.}~\bibnamefont {Koivisto}},\ and\ \bibinfo {author} {\bibfnamefont {N.~J.}\ \bibnamefont {Nunes}},\ }\bibfield  {title} {\bibinfo {title} {{Testing $F(Q)$ gravity with redshift space distortions}},\ }\href {https://doi.org/10.1016/j.dark.2020.100616} {\bibfield  {journal} {\bibinfo  {journal} {Phys. Dark Univ.}\ }\textbf {\bibinfo {volume} {30}},\ \bibinfo {pages} {100616} (\bibinfo {year} {2020})},\ \Eprint {https://arxiv.org/abs/2004.07867} {arXiv:2004.07867 [gr-qc]} \BibitemShut {NoStop}%
\bibitem [{\citenamefont {Dimakis}\ \emph {et~al.}(2022)\citenamefont {Dimakis}, \citenamefont {Paliathanasis}, \citenamefont {Roumeliotis},\ and\ \citenamefont {Christodoulakis}}]{Dimakis:2022rkd}%
  \BibitemOpen
  \bibfield  {author} {\bibinfo {author} {\bibfnamefont {N.}~\bibnamefont {Dimakis}}, \bibinfo {author} {\bibfnamefont {A.}~\bibnamefont {Paliathanasis}}, \bibinfo {author} {\bibfnamefont {M.}~\bibnamefont {Roumeliotis}},\ and\ \bibinfo {author} {\bibfnamefont {T.}~\bibnamefont {Christodoulakis}},\ }\bibfield  {title} {\bibinfo {title} {{FLRW solutions in f(Q) theory: The effect of using different connections}},\ }\href {https://doi.org/10.1103/PhysRevD.106.043509} {\bibfield  {journal} {\bibinfo  {journal} {Phys. Rev. D}\ }\textbf {\bibinfo {volume} {106}},\ \bibinfo {pages} {043509} (\bibinfo {year} {2022})},\ \Eprint {https://arxiv.org/abs/2205.04680} {arXiv:2205.04680 [gr-qc]} \BibitemShut {NoStop}%
\bibitem [{\citenamefont {Khyllep}\ \emph {et~al.}(2023)\citenamefont {Khyllep}, \citenamefont {Dutta}, \citenamefont {Saridakis},\ and\ \citenamefont {Yesmakhanova}}]{Khyllep:2022spx}%
  \BibitemOpen
  \bibfield  {author} {\bibinfo {author} {\bibfnamefont {W.}~\bibnamefont {Khyllep}}, \bibinfo {author} {\bibfnamefont {J.}~\bibnamefont {Dutta}}, \bibinfo {author} {\bibfnamefont {E.~N.}\ \bibnamefont {Saridakis}},\ and\ \bibinfo {author} {\bibfnamefont {K.}~\bibnamefont {Yesmakhanova}},\ }\bibfield  {title} {\bibinfo {title} {{Cosmology in f(Q) gravity: A unified dynamical systems analysis of the background and perturbations}},\ }\href {https://doi.org/10.1103/PhysRevD.107.044022} {\bibfield  {journal} {\bibinfo  {journal} {Phys. Rev. D}\ }\textbf {\bibinfo {volume} {107}},\ \bibinfo {pages} {044022} (\bibinfo {year} {2023})},\ \Eprint {https://arxiv.org/abs/2207.02610} {arXiv:2207.02610 [gr-qc]} \BibitemShut {NoStop}%
\bibitem [{\citenamefont {Subramaniam}\ \emph {et~al.}(2023)\citenamefont {Subramaniam}, \citenamefont {De}, \citenamefont {Loo},\ and\ \citenamefont {Goh}}]{Subramaniam:2023okn}%
  \BibitemOpen
  \bibfield  {author} {\bibinfo {author} {\bibfnamefont {G.}~\bibnamefont {Subramaniam}}, \bibinfo {author} {\bibfnamefont {A.}~\bibnamefont {De}}, \bibinfo {author} {\bibfnamefont {T.-H.}\ \bibnamefont {Loo}},\ and\ \bibinfo {author} {\bibfnamefont {Y.~K.}\ \bibnamefont {Goh}},\ }\bibfield  {title} {\bibinfo {title} {{How Different Connections in Flat FLRW Geometry Impact Energy Conditions in f(Q)$f(Q)$ Theory?}},\ }\href {https://doi.org/10.1002/prop.202300038} {\bibfield  {journal} {\bibinfo  {journal} {Fortsch. Phys.}\ }\textbf {\bibinfo {volume} {71}},\ \bibinfo {pages} {2300038} (\bibinfo {year} {2023})},\ \Eprint {https://arxiv.org/abs/2304.02300} {arXiv:2304.02300 [gr-qc]} \BibitemShut {NoStop}%
\bibitem [{\citenamefont {Bajardi}\ and\ \citenamefont {Capozziello}(2023)}]{Bajardi:2023vcc}%
  \BibitemOpen
  \bibfield  {author} {\bibinfo {author} {\bibfnamefont {F.}~\bibnamefont {Bajardi}}\ and\ \bibinfo {author} {\bibfnamefont {S.}~\bibnamefont {Capozziello}},\ }\bibfield  {title} {\bibinfo {title} {{Minisuperspace quantum cosmology in f(Q) gravity}},\ }\href {https://doi.org/10.1140/epjc/s10052-023-11703-8} {\bibfield  {journal} {\bibinfo  {journal} {Eur. Phys. J. C}\ }\textbf {\bibinfo {volume} {83}},\ \bibinfo {pages} {531} (\bibinfo {year} {2023})},\ \Eprint {https://arxiv.org/abs/2305.00318} {arXiv:2305.00318 [gr-qc]} \BibitemShut {NoStop}%
\bibitem [{\citenamefont {Heisenberg}(2024)}]{Heisenberg:2023lru}%
  \BibitemOpen
  \bibfield  {author} {\bibinfo {author} {\bibfnamefont {L.}~\bibnamefont {Heisenberg}},\ }\bibfield  {title} {\bibinfo {title} {{Review on f(Q) gravity}},\ }\href {https://doi.org/10.1016/j.physrep.2024.02.001} {\bibfield  {journal} {\bibinfo  {journal} {Phys. Rept.}\ }\textbf {\bibinfo {volume} {1066}},\ \bibinfo {pages} {1} (\bibinfo {year} {2024})},\ \Eprint {https://arxiv.org/abs/2309.15958} {arXiv:2309.15958 [gr-qc]} \BibitemShut {NoStop}%
\bibitem [{\citenamefont {Guzm{\'a}n}\ \emph {et~al.}(2024)\citenamefont {Guzm{\'a}n}, \citenamefont {J{\"a}rv},\ and\ \citenamefont {Pati}}]{Guzman:2024cwa}%
  \BibitemOpen
  \bibfield  {author} {\bibinfo {author} {\bibfnamefont {M.-J.}\ \bibnamefont {Guzm{\'a}n}}, \bibinfo {author} {\bibfnamefont {L.}~\bibnamefont {J{\"a}rv}},\ and\ \bibinfo {author} {\bibfnamefont {L.}~\bibnamefont {Pati}},\ }\bibfield  {title} {\bibinfo {title} {{Exploring the stability of f(Q) cosmology near general relativity limit with different connections}},\ }\href {https://doi.org/10.1103/PhysRevD.110.124013} {\bibfield  {journal} {\bibinfo  {journal} {Phys. Rev. D}\ }\textbf {\bibinfo {volume} {110}},\ \bibinfo {pages} {124013} (\bibinfo {year} {2024})},\ \Eprint {https://arxiv.org/abs/2406.11621} {arXiv:2406.11621 [gr-qc]} \BibitemShut {NoStop}%
\bibitem [{\citenamefont {D'Ambrosio}\ \emph {et~al.}(2022)\citenamefont {D'Ambrosio}, \citenamefont {Fell}, \citenamefont {Heisenberg},\ and\ \citenamefont {Kuhn}}]{DAmbrosio:2021zpm}%
  \BibitemOpen
  \bibfield  {author} {\bibinfo {author} {\bibfnamefont {F.}~\bibnamefont {D'Ambrosio}}, \bibinfo {author} {\bibfnamefont {S.~D.~B.}\ \bibnamefont {Fell}}, \bibinfo {author} {\bibfnamefont {L.}~\bibnamefont {Heisenberg}},\ and\ \bibinfo {author} {\bibfnamefont {S.}~\bibnamefont {Kuhn}},\ }\bibfield  {title} {\bibinfo {title} {{Black holes in f(Q) gravity}},\ }\href {https://doi.org/10.1103/PhysRevD.105.024042} {\bibfield  {journal} {\bibinfo  {journal} {Phys. Rev. D}\ }\textbf {\bibinfo {volume} {105}},\ \bibinfo {pages} {024042} (\bibinfo {year} {2022})},\ \Eprint {https://arxiv.org/abs/2109.03174} {arXiv:2109.03174 [gr-qc]} \BibitemShut {NoStop}%
\bibitem [{\citenamefont {Lin}\ and\ \citenamefont {Zhai}(2021)}]{Lin:2021uqa}%
  \BibitemOpen
  \bibfield  {author} {\bibinfo {author} {\bibfnamefont {R.-H.}\ \bibnamefont {Lin}}\ and\ \bibinfo {author} {\bibfnamefont {X.-H.}\ \bibnamefont {Zhai}},\ }\bibfield  {title} {\bibinfo {title} {{Spherically symmetric configuration in $f(Q)$ gravity}},\ }\href {https://doi.org/10.1103/PhysRevD.103.124001} {\bibfield  {journal} {\bibinfo  {journal} {Phys. Rev. D}\ }\textbf {\bibinfo {volume} {103}},\ \bibinfo {pages} {124001} (\bibinfo {year} {2021})},\ \bibinfo {note} {[Erratum: Phys.Rev.D 106, 069902 (2022)]},\ \Eprint {https://arxiv.org/abs/2105.01484} {arXiv:2105.01484 [gr-qc]} \BibitemShut {NoStop}%
\bibitem [{\citenamefont {Wang}\ \emph {et~al.}(2022)\citenamefont {Wang}, \citenamefont {Chen},\ and\ \citenamefont {Katsuragawa}}]{Wang:2021zaz}%
  \BibitemOpen
  \bibfield  {author} {\bibinfo {author} {\bibfnamefont {W.}~\bibnamefont {Wang}}, \bibinfo {author} {\bibfnamefont {H.}~\bibnamefont {Chen}},\ and\ \bibinfo {author} {\bibfnamefont {T.}~\bibnamefont {Katsuragawa}},\ }\bibfield  {title} {\bibinfo {title} {{Static and spherically symmetric solutions in f(Q) gravity}},\ }\href {https://doi.org/10.1103/PhysRevD.105.024060} {\bibfield  {journal} {\bibinfo  {journal} {Phys. Rev. D}\ }\textbf {\bibinfo {volume} {105}},\ \bibinfo {pages} {024060} (\bibinfo {year} {2022})},\ \Eprint {https://arxiv.org/abs/2110.13565} {arXiv:2110.13565 [gr-qc]} \BibitemShut {NoStop}%
\bibitem [{\citenamefont {Calz{\'a}}\ and\ \citenamefont {Sebastiani}(2023)}]{Calza:2022mwt}%
  \BibitemOpen
  \bibfield  {author} {\bibinfo {author} {\bibfnamefont {M.}~\bibnamefont {Calz{\'a}}}\ and\ \bibinfo {author} {\bibfnamefont {L.}~\bibnamefont {Sebastiani}},\ }\bibfield  {title} {\bibinfo {title} {{A class of static spherically symmetric solutions in f(Q)-gravity}},\ }\href {https://doi.org/10.1140/epjc/s10052-023-11393-2} {\bibfield  {journal} {\bibinfo  {journal} {Eur. Phys. J. C}\ }\textbf {\bibinfo {volume} {83}},\ \bibinfo {pages} {247} (\bibinfo {year} {2023})},\ \Eprint {https://arxiv.org/abs/2208.13033} {arXiv:2208.13033 [gr-qc]} \BibitemShut {NoStop}%
\bibitem [{\citenamefont {Bahamonde}\ and\ \citenamefont {J{\"a}rv}(2022)}]{Bahamonde:2022zgj}%
  \BibitemOpen
  \bibfield  {author} {\bibinfo {author} {\bibfnamefont {S.}~\bibnamefont {Bahamonde}}\ and\ \bibinfo {author} {\bibfnamefont {L.}~\bibnamefont {J{\"a}rv}},\ }\bibfield  {title} {\bibinfo {title} {{Coincident gauge for static spherical field configurations in symmetric teleparallel gravity}},\ }\href {https://doi.org/10.1140/epjc/s10052-022-10922-9} {\bibfield  {journal} {\bibinfo  {journal} {Eur. Phys. J. C}\ }\textbf {\bibinfo {volume} {82}},\ \bibinfo {pages} {963} (\bibinfo {year} {2022})},\ \Eprint {https://arxiv.org/abs/2208.01872} {arXiv:2208.01872 [gr-qc]} \BibitemShut {NoStop}%
\bibitem [{\citenamefont {Hohmann}\ and\ \citenamefont {Karanasou}(2025)}]{Hohmann:2024phz}%
  \BibitemOpen
  \bibfield  {author} {\bibinfo {author} {\bibfnamefont {M.}~\bibnamefont {Hohmann}}\ and\ \bibinfo {author} {\bibfnamefont {V.}~\bibnamefont {Karanasou}},\ }\bibfield  {title} {\bibinfo {title} {{Symmetric teleparallel connection and spherical solutions in symmetric teleparallel gravity}},\ }\href {https://doi.org/10.1103/PhysRevD.111.064057} {\bibfield  {journal} {\bibinfo  {journal} {Phys. Rev. D}\ }\textbf {\bibinfo {volume} {111}},\ \bibinfo {pages} {064057} (\bibinfo {year} {2025})},\ \Eprint {https://arxiv.org/abs/2412.11730} {arXiv:2412.11730 [gr-qc]} \BibitemShut {NoStop}%
\bibitem [{\citenamefont {Zhao}(2022)}]{Zhao:2021zab}%
  \BibitemOpen
  \bibfield  {author} {\bibinfo {author} {\bibfnamefont {D.}~\bibnamefont {Zhao}},\ }\bibfield  {title} {\bibinfo {title} {{Covariant formulation of f(Q) theory}},\ }\href {https://doi.org/10.1140/epjc/s10052-022-10266-4} {\bibfield  {journal} {\bibinfo  {journal} {Eur. Phys. J. C}\ }\textbf {\bibinfo {volume} {82}},\ \bibinfo {pages} {303} (\bibinfo {year} {2022})},\ \Eprint {https://arxiv.org/abs/2104.02483} {arXiv:2104.02483 [gr-qc]} \BibitemShut {NoStop}%
\bibitem [{\citenamefont {Ipser}\ and\ \citenamefont {Sikivie}(1984)}]{Ipser:1984db}%
  \BibitemOpen
  \bibfield  {author} {\bibinfo {author} {\bibfnamefont {J.}~\bibnamefont {Ipser}}\ and\ \bibinfo {author} {\bibfnamefont {P.}~\bibnamefont {Sikivie}},\ }\bibfield  {title} {\bibinfo {title} {{The Gravitationally Repulsive Domain Wall}},\ }\href {https://doi.org/10.1103/PhysRevD.30.712} {\bibfield  {journal} {\bibinfo  {journal} {Phys. Rev. D}\ }\textbf {\bibinfo {volume} {30}},\ \bibinfo {pages} {712} (\bibinfo {year} {1984})}\BibitemShut {NoStop}%
\bibitem [{\citenamefont {Boulanger}\ \emph {et~al.}(2006)\citenamefont {Boulanger}, \citenamefont {Spindel},\ and\ \citenamefont {Buisseret}}]{PhysRevD.74.125014}%
  \BibitemOpen
  \bibfield  {author} {\bibinfo {author} {\bibfnamefont {N.}~\bibnamefont {Boulanger}}, \bibinfo {author} {\bibfnamefont {P.}~\bibnamefont {Spindel}},\ and\ \bibinfo {author} {\bibfnamefont {F.}~\bibnamefont {Buisseret}},\ }\bibfield  {title} {\bibinfo {title} {Bound states of dirac particles in gravitational fields},\ }\href {https://doi.org/10.1103/PhysRevD.74.125014} {\bibfield  {journal} {\bibinfo  {journal} {Phys. Rev. D}\ }\textbf {\bibinfo {volume} {74}},\ \bibinfo {pages} {125014} (\bibinfo {year} {2006})}\BibitemShut {NoStop}%
\bibitem [{\citenamefont {Lanosa}\ and\ \citenamefont {Santill{\'o}n}(2024)}]{Lanosa:2023mox}%
  \BibitemOpen
  \bibfield  {author} {\bibinfo {author} {\bibfnamefont {L.}~\bibnamefont {Lanosa}}\ and\ \bibinfo {author} {\bibfnamefont {O.~P.}\ \bibnamefont {Santill{\'o}n}},\ }\bibfield  {title} {\bibinfo {title} {{Peculiarities for domain walls in Taub coordinates}},\ }\href {https://doi.org/10.1140/epjc/s10052-024-12384-7} {\bibfield  {journal} {\bibinfo  {journal} {Eur. Phys. J. C}\ }\textbf {\bibinfo {volume} {84}},\ \bibinfo {pages} {33} (\bibinfo {year} {2024})},\ \Eprint {https://arxiv.org/abs/2310.10968} {arXiv:2310.10968 [gr-qc]} \BibitemShut {NoStop}%
\bibitem [{\citenamefont {Taub}(1951)}]{Taub:1951ez}%
  \BibitemOpen
  \bibfield  {author} {\bibinfo {author} {\bibfnamefont {A.~H.}\ \bibnamefont {Taub}},\ }\bibfield  {title} {\bibinfo {title} {{Empty space-times admitting a three parameter group of motions}},\ }\href {https://doi.org/10.2307/1969567} {\bibfield  {journal} {\bibinfo  {journal} {Annals Math.}\ }\textbf {\bibinfo {volume} {53}},\ \bibinfo {pages} {472} (\bibinfo {year} {1951})}\BibitemShut {NoStop}%
\bibitem [{\citenamefont {Aichelburg}(1970)}]{Aichelburg:1970}%
  \BibitemOpen
  \bibfield  {author} {\bibinfo {author} {\bibfnamefont {P.~C.}\ \bibnamefont {Aichelburg}},\ }\bibfield  {title} {\bibinfo {title} {High symmetry fields and the homogeneous field in general relativity},\ }\href {https://api.semanticscholar.org/CorpusID:121800211} {\bibfield  {journal} {\bibinfo  {journal} {Journal of Mathematical Physics}\ }\textbf {\bibinfo {volume} {11}},\ \bibinfo {pages} {1330} (\bibinfo {year} {1970})}\BibitemShut {NoStop}%
\bibitem [{\citenamefont {Bedran}\ \emph {et~al.}(1997)\citenamefont {Bedran}, \citenamefont {Calvao}, \citenamefont {Soares},\ and\ \citenamefont {Paiva}}]{Bedran:1997su}%
  \BibitemOpen
  \bibfield  {author} {\bibinfo {author} {\bibfnamefont {M.~L.}\ \bibnamefont {Bedran}}, \bibinfo {author} {\bibfnamefont {M.~O.}\ \bibnamefont {Calvao}}, \bibinfo {author} {\bibfnamefont {I.~D.}\ \bibnamefont {Soares}},\ and\ \bibinfo {author} {\bibfnamefont {F.~M.}\ \bibnamefont {Paiva}},\ }\bibfield  {title} {\bibinfo {title} {{Taub's plane symmetric vacuum space-time revisited}},\ }\href {https://doi.org/10.1103/PhysRevD.55.3431} {\bibfield  {journal} {\bibinfo  {journal} {Phys. Rev. D}\ }\textbf {\bibinfo {volume} {55}},\ \bibinfo {pages} {3431} (\bibinfo {year} {1997})},\ \Eprint {https://arxiv.org/abs/gr-qc/9608058} {arXiv:gr-qc/9608058} \BibitemShut {NoStop}%
\bibitem [{\citenamefont {Groen}\ and\ \citenamefont {Soleng}(1992)}]{Groen:1992sm}%
  \BibitemOpen
  \bibfield  {author} {\bibinfo {author} {\bibfnamefont {O.}~\bibnamefont {Groen}}\ and\ \bibinfo {author} {\bibfnamefont {H.~H.}\ \bibnamefont {Soleng}},\ }\bibfield  {title} {\bibinfo {title} {{Static plane symmetric space-time with a conformally coupled massless scalar field}},\ }\href {https://doi.org/10.1016/0375-9601(92)90033-I} {\bibfield  {journal} {\bibinfo  {journal} {Phys. Lett. A}\ }\textbf {\bibinfo {volume} {165}},\ \bibinfo {pages} {191} (\bibinfo {year} {1992})}\BibitemShut {NoStop}%
\bibitem [{\citenamefont {Jensen}\ and\ \citenamefont {Kucera}(1994)}]{Jensen:1994zf}%
  \BibitemOpen
  \bibfield  {author} {\bibinfo {author} {\bibfnamefont {B.}~\bibnamefont {Jensen}}\ and\ \bibinfo {author} {\bibfnamefont {J.}~\bibnamefont {Kucera}},\ }\bibfield  {title} {\bibinfo {title} {{A Reinterpretation of the Taub singularity}},\ }\href {https://doi.org/10.1016/0375-9601(94)90081-7} {\bibfield  {journal} {\bibinfo  {journal} {Phys. Lett. A}\ }\textbf {\bibinfo {volume} {195}},\ \bibinfo {pages} {111} (\bibinfo {year} {1994})},\ \Eprint {https://arxiv.org/abs/gr-qc/9406035} {arXiv:gr-qc/9406035} \BibitemShut {NoStop}%
\bibitem [{\citenamefont {Gamboa~Saravi}(2008)}]{GamboaSaravi:2007se}%
  \BibitemOpen
  \bibfield  {author} {\bibinfo {author} {\bibfnamefont {R.~E.}\ \bibnamefont {Gamboa~Saravi}},\ }\bibfield  {title} {\bibinfo {title} {{Static plane symmetric relativistic fluids and empty repelling singular boundaries}},\ }\href {https://doi.org/10.1088/0264-9381/25/4/045005} {\bibfield  {journal} {\bibinfo  {journal} {Class. Quant. Grav.}\ }\textbf {\bibinfo {volume} {25}},\ \bibinfo {pages} {045005} (\bibinfo {year} {2008})},\ \Eprint {https://arxiv.org/abs/0712.2831} {arXiv:0712.2831 [gr-qc]} \BibitemShut {NoStop}%
\bibitem [{\citenamefont {da~Silva}\ \emph {et~al.}(1998)\citenamefont {da~Silva}, \citenamefont {Wang},\ and\ \citenamefont {Santos}}]{daSilva:1997uf}%
  \BibitemOpen
  \bibfield  {author} {\bibinfo {author} {\bibfnamefont {M.~F.~A.}\ \bibnamefont {da~Silva}}, \bibinfo {author} {\bibfnamefont {A.}~\bibnamefont {Wang}},\ and\ \bibinfo {author} {\bibfnamefont {N.~O.}\ \bibnamefont {Santos}},\ }\bibfield  {title} {\bibinfo {title} {{On the sources of static plane symmetric vacuum space-times}},\ }\href {https://doi.org/10.1016/S0375-9601(98)00355-7} {\bibfield  {journal} {\bibinfo  {journal} {Phys. Lett. A}\ }\textbf {\bibinfo {volume} {244}},\ \bibinfo {pages} {462} (\bibinfo {year} {1998})},\ \Eprint {https://arxiv.org/abs/gr-qc/9706071} {arXiv:gr-qc/9706071} \BibitemShut {NoStop}%
\bibitem [{\citenamefont {Gamboa~Saravi}(2009)}]{GamboaSaravi:2009sw}%
  \BibitemOpen
  \bibfield  {author} {\bibinfo {author} {\bibfnamefont {R.~E.}\ \bibnamefont {Gamboa~Saravi}},\ }\bibfield  {title} {\bibinfo {title} {{The Gravitational Field of a Plane Slab}},\ }\href {https://doi.org/10.1142/S0217751X09046096} {\bibfield  {journal} {\bibinfo  {journal} {Int. J. Mod. Phys. A}\ }\textbf {\bibinfo {volume} {24}},\ \bibinfo {pages} {5381} (\bibinfo {year} {2009})},\ \Eprint {https://arxiv.org/abs/0905.0896} {arXiv:0905.0896 [gr-qc]} \BibitemShut {NoStop}%
\bibitem [{\citenamefont {Fulling}\ \emph {et~al.}(2015)\citenamefont {Fulling}, \citenamefont {Bouas},\ and\ \citenamefont {Carter}}]{Fulling:2015uva}%
  \BibitemOpen
  \bibfield  {author} {\bibinfo {author} {\bibfnamefont {S.~A.}\ \bibnamefont {Fulling}}, \bibinfo {author} {\bibfnamefont {J.~D.}\ \bibnamefont {Bouas}},\ and\ \bibinfo {author} {\bibfnamefont {H.~B.}\ \bibnamefont {Carter}},\ }\bibfield  {title} {\bibinfo {title} {{The gravitational field of an infinite flat slab}},\ }\href {https://doi.org/10.1088/0031-8949/90/8/088006} {\bibfield  {journal} {\bibinfo  {journal} {Phys. Scripta}\ }\textbf {\bibinfo {volume} {90}},\ \bibinfo {pages} {088006} (\bibinfo {year} {2015})}\BibitemShut {NoStop}%
\bibitem [{\citenamefont {Acu\~na}\ and\ \citenamefont {Perico~Esguerra}(2015)}]{Acuna:2015tsa}%
  \BibitemOpen
  \bibfield  {author} {\bibinfo {author} {\bibfnamefont {J.~T.}\ \bibnamefont {Acu\~na}}\ and\ \bibinfo {author} {\bibfnamefont {J.}~\bibnamefont {Perico~Esguerra}},\ }\bibfield  {title} {\bibinfo {title} {{Dynamics of a planar thin shell at a Taub\textendash{}FRW junction}},\ }\href {https://doi.org/10.1142/S0218271816500012} {\bibfield  {journal} {\bibinfo  {journal} {Int. J. Mod. Phys. D}\ }\textbf {\bibinfo {volume} {25}},\ \bibinfo {pages} {1650001} (\bibinfo {year} {2015})},\ \Eprint {https://arxiv.org/abs/1509.07965} {arXiv:1509.07965 [gr-qc]} \BibitemShut {NoStop}%
\bibitem [{\citenamefont {Kamenshchik}\ and\ \citenamefont {Vardanyan}(2020)}]{Kamenshchik:2020div}%
  \BibitemOpen
  \bibfield  {author} {\bibinfo {author} {\bibfnamefont {A.~Y.}\ \bibnamefont {Kamenshchik}}\ and\ \bibinfo {author} {\bibfnamefont {T.}~\bibnamefont {Vardanyan}},\ }\bibfield  {title} {\bibinfo {title} {{Spatial Kasner Solution and an Infinite Slab with a Constant Energy Density}},\ }\href {https://doi.org/10.1134/S0021364020060016} {\bibfield  {journal} {\bibinfo  {journal} {JETP Lett.}\ }\textbf {\bibinfo {volume} {111}},\ \bibinfo {pages} {306} (\bibinfo {year} {2020})},\ \Eprint {https://arxiv.org/abs/2002.12916} {arXiv:2002.12916 [gr-qc]} \BibitemShut {NoStop}%
\bibitem [{\citenamefont {Avagyan}\ \emph {et~al.}(2024)\citenamefont {Avagyan}, \citenamefont {Petrosyan}, \citenamefont {Saharian},\ and\ \citenamefont {Harutyunyan}}]{Avagyan:2024oia}%
  \BibitemOpen
  \bibfield  {author} {\bibinfo {author} {\bibfnamefont {R.~M.}\ \bibnamefont {Avagyan}}, \bibinfo {author} {\bibfnamefont {T.~A.}\ \bibnamefont {Petrosyan}}, \bibinfo {author} {\bibfnamefont {A.~A.}\ \bibnamefont {Saharian}},\ and\ \bibinfo {author} {\bibfnamefont {G.~H.}\ \bibnamefont {Harutyunyan}},\ }\bibfield  {title} {\bibinfo {title} {{Plane Symmetric Gravitational Fields in (D+1)-dimensional General Relativity}},\ }\href {https://doi.org/10.1007/s10511-024-09840-3} {\bibfield  {journal} {\bibinfo  {journal} {Astrophysics}\ }\textbf {\bibinfo {volume} {67}},\ \bibinfo {pages} {405} (\bibinfo {year} {2024})},\ \Eprint {https://arxiv.org/abs/2406.13758} {arXiv:2406.13758 [gr-qc]} \BibitemShut {NoStop}%
\bibitem [{\citenamefont {Zhang}\ \emph {et~al.}(2008)\citenamefont {Zhang}, \citenamefont {Noh},\ and\ \citenamefont {Zhu}}]{Zhang:2008ss}%
  \BibitemOpen
  \bibfield  {author} {\bibinfo {author} {\bibfnamefont {H.-s.}\ \bibnamefont {Zhang}}, \bibinfo {author} {\bibfnamefont {H.}~\bibnamefont {Noh}},\ and\ \bibinfo {author} {\bibfnamefont {Z.-H.}\ \bibnamefont {Zhu}},\ }\bibfield  {title} {\bibinfo {title} {{A New class of plane symmetric solution}},\ }\href {https://doi.org/10.1016/j.physletb.2008.04.022} {\bibfield  {journal} {\bibinfo  {journal} {Phys. Lett. B}\ }\textbf {\bibinfo {volume} {663}},\ \bibinfo {pages} {291} (\bibinfo {year} {2008})},\ \Eprint {https://arxiv.org/abs/0804.2931} {arXiv:0804.2931 [gr-qc]} \BibitemShut {NoStop}%
\bibitem [{\citenamefont {Zhang}\ and\ \citenamefont {Noh}(2009{\natexlab{a}})}]{Zhang:2009hx}%
  \BibitemOpen
  \bibfield  {author} {\bibinfo {author} {\bibfnamefont {H.}~\bibnamefont {Zhang}}\ and\ \bibinfo {author} {\bibfnamefont {H.}~\bibnamefont {Noh}},\ }\bibfield  {title} {\bibinfo {title} {{Some properties of a new class of plane symmetric solution}},\ }\href {https://doi.org/10.1016/j.physletb.2008.11.015} {\bibfield  {journal} {\bibinfo  {journal} {Phys. Lett. B}\ }\textbf {\bibinfo {volume} {670}},\ \bibinfo {pages} {271} (\bibinfo {year} {2009}{\natexlab{a}})},\ \Eprint {https://arxiv.org/abs/0904.0063} {arXiv:0904.0063 [gr-qc]} \BibitemShut {NoStop}%
\bibitem [{\citenamefont {Zhang}\ and\ \citenamefont {Noh}(2009{\natexlab{b}})}]{Zhang:2009hz}%
  \BibitemOpen
  \bibfield  {author} {\bibinfo {author} {\bibfnamefont {H.}~\bibnamefont {Zhang}}\ and\ \bibinfo {author} {\bibfnamefont {H.}~\bibnamefont {Noh}},\ }\bibfield  {title} {\bibinfo {title} {{N-dimensional plane symmetric solution with perfect fluid source}},\ }\href {https://doi.org/10.1016/j.physletb.2008.12.057} {\bibfield  {journal} {\bibinfo  {journal} {Phys. Lett. B}\ }\textbf {\bibinfo {volume} {671}},\ \bibinfo {pages} {428} (\bibinfo {year} {2009}{\natexlab{b}})},\ \Eprint {https://arxiv.org/abs/0904.0065} {arXiv:0904.0065 [gr-qc]} \BibitemShut {NoStop}%
\bibitem [{\citenamefont {Zhang}\ and\ \citenamefont {Li}(2011)}]{Zhang:2011zzq}%
  \BibitemOpen
  \bibfield  {author} {\bibinfo {author} {\bibfnamefont {H.}~\bibnamefont {Zhang}}\ and\ \bibinfo {author} {\bibfnamefont {X.-Z.}\ \bibnamefont {Li}},\ }\bibfield  {title} {\bibinfo {title} {{Some properties of the n-dimensional plane symmetric solution}},\ }\href {https://doi.org/10.1016/j.physletb.2011.04.058} {\bibfield  {journal} {\bibinfo  {journal} {Phys. Lett. B}\ }\textbf {\bibinfo {volume} {700}},\ \bibinfo {pages} {97} (\bibinfo {year} {2011})}\BibitemShut {NoStop}%
\bibitem [{\citenamefont {Sharif}\ and\ \citenamefont {Shamir}(2010)}]{Sharif:2009uc}%
  \BibitemOpen
  \bibfield  {author} {\bibinfo {author} {\bibfnamefont {M.}~\bibnamefont {Sharif}}\ and\ \bibinfo {author} {\bibfnamefont {M.~F.}\ \bibnamefont {Shamir}},\ }\bibfield  {title} {\bibinfo {title} {{Plane Symmetric Solutions in f(R) Gravity}},\ }\href {https://doi.org/10.1142/S0217732310032536} {\bibfield  {journal} {\bibinfo  {journal} {Mod. Phys. Lett. A}\ }\textbf {\bibinfo {volume} {25}},\ \bibinfo {pages} {1281} (\bibinfo {year} {2010})},\ \Eprint {https://arxiv.org/abs/0912.1393} {arXiv:0912.1393 [gr-qc]} \BibitemShut {NoStop}%
\bibitem [{\citenamefont {Yavari}(2013)}]{Yavari:2013fga}%
  \BibitemOpen
  \bibfield  {author} {\bibinfo {author} {\bibfnamefont {M.}~\bibnamefont {Yavari}},\ }\bibfield  {title} {\bibinfo {title} {{The plane symmetric vacuum solutions of modified field equations in metric $f(R)$ gravity}},\ }\href {https://doi.org/10.1007/s10509-013-1565-4} {\bibfield  {journal} {\bibinfo  {journal} {Astrophys. Space Sci.}\ }\textbf {\bibinfo {volume} {348}},\ \bibinfo {pages} {293} (\bibinfo {year} {2013})}\BibitemShut {NoStop}%
\bibitem [{\citenamefont {\"Oz}\ and\ \citenamefont {Bamba}(2022)}]{Oz:2021jnw}%
  \BibitemOpen
  \bibfield  {author} {\bibinfo {author} {\bibfnamefont {I.~B.}\ \bibnamefont {\"Oz}}\ and\ \bibinfo {author} {\bibfnamefont {K.}~\bibnamefont {Bamba}},\ }\bibfield  {title} {\bibinfo {title} {{Cylindrically symmetric and plane-symmetric solutions in f(R) theory via Noether symmetries}},\ }\href {https://doi.org/10.1140/epjc/s10052-022-10316-x} {\bibfield  {journal} {\bibinfo  {journal} {Eur. Phys. J. C}\ }\textbf {\bibinfo {volume} {82}},\ \bibinfo {pages} {349} (\bibinfo {year} {2022})},\ \Eprint {https://arxiv.org/abs/2111.05750} {arXiv:2111.05750 [gr-qc]} \BibitemShut {NoStop}%
\bibitem [{\citenamefont {Agrawal}\ and\ \citenamefont {Pawar}(2017)}]{Agrawal:2017wvt}%
  \BibitemOpen
  \bibfield  {author} {\bibinfo {author} {\bibfnamefont {P.~K.}\ \bibnamefont {Agrawal}}\ and\ \bibinfo {author} {\bibfnamefont {D.~D.}\ \bibnamefont {Pawar}},\ }\bibfield  {title} {\bibinfo {title} {{Plane Symmetric Cosmological Model with Quark and Strange Quark Matter in f(R,T) Theory of Gravity}},\ }\href {https://doi.org/10.1007/s12036-016-9420-y} {\bibfield  {journal} {\bibinfo  {journal} {J. Astrophys. Astron.}\ }\textbf {\bibinfo {volume} {38}},\ \bibinfo {pages} {2} (\bibinfo {year} {2017})},\ \Eprint {https://arxiv.org/abs/1702.02349} {arXiv:1702.02349 [gr-qc]} \BibitemShut {NoStop}%
\bibitem [{\citenamefont {Mehmood}\ \emph {et~al.}(2022)\citenamefont {Mehmood}, \citenamefont {Hussain}, \citenamefont {Bokhari}, \citenamefont {Ramzan}, \citenamefont {Faryad},\ and\ \citenamefont {Hussain}}]{Mehmood:2022zcg}%
  \BibitemOpen
  \bibfield  {author} {\bibinfo {author} {\bibfnamefont {A.~B.}\ \bibnamefont {Mehmood}}, \bibinfo {author} {\bibfnamefont {F.}~\bibnamefont {Hussain}}, \bibinfo {author} {\bibfnamefont {A.~H.}\ \bibnamefont {Bokhari}}, \bibinfo {author} {\bibfnamefont {M.}~\bibnamefont {Ramzan}}, \bibinfo {author} {\bibfnamefont {M.}~\bibnamefont {Faryad}},\ and\ \bibinfo {author} {\bibfnamefont {T.}~\bibnamefont {Hussain}},\ }\bibfield  {title} {\bibinfo {title} {{On some non-static plane symmetric perfect fluid solutions in f(R,T) gravity}},\ }\href {https://doi.org/10.1016/j.rinp.2022.105676} {\bibfield  {journal} {\bibinfo  {journal} {Results Phys.}\ }\textbf {\bibinfo {volume} {39}},\ \bibinfo {pages} {105676} (\bibinfo {year} {2022})}\BibitemShut {NoStop}%
\bibitem [{\citenamefont {Singh}\ \emph {et~al.}(2023)\citenamefont {Singh}, \citenamefont {Jokweni},\ and\ \citenamefont {Beesham}}]{Singh:2023mgr}%
  \BibitemOpen
  \bibfield  {author} {\bibinfo {author} {\bibfnamefont {V.}~\bibnamefont {Singh}}, \bibinfo {author} {\bibfnamefont {S.}~\bibnamefont {Jokweni}},\ and\ \bibinfo {author} {\bibfnamefont {A.}~\bibnamefont {Beesham}},\ }\bibfield  {title} {\bibinfo {title} {{Plane Symmetric Cosmological Model with Strange Quark Matter in $f(R,T)$ Gravity}},\ }\href {https://doi.org/10.3390/universe9090408} {\bibfield  {journal} {\bibinfo  {journal} {Universe}\ }\textbf {\bibinfo {volume} {9}},\ \bibinfo {pages} {408} (\bibinfo {year} {2023})}\BibitemShut {NoStop}%
\bibitem [{\citenamefont {Dalal}\ \emph {et~al.}(2024)\citenamefont {Dalal}, \citenamefont {Singh},\ and\ \citenamefont {Kumar}}]{Dalal:2024qkm}%
  \BibitemOpen
  \bibfield  {author} {\bibinfo {author} {\bibfnamefont {P.}~\bibnamefont {Dalal}}, \bibinfo {author} {\bibfnamefont {K.}~\bibnamefont {Singh}},\ and\ \bibinfo {author} {\bibfnamefont {S.}~\bibnamefont {Kumar}},\ }\bibfield  {title} {\bibinfo {title} {{Non-static plane symmetric perfect fluid solutions and Killing symmetries in f(R, T) gravity}},\ }\href {https://doi.org/10.1088/1572-9494/ad08ab} {\bibfield  {journal} {\bibinfo  {journal} {Commun. Theor. Phys.}\ }\textbf {\bibinfo {volume} {76}},\ \bibinfo {pages} {025406} (\bibinfo {year} {2024})}\BibitemShut {NoStop}%
\bibitem [{\citenamefont {Sharif}\ and\ \citenamefont {Azeem}(2012)}]{Sharif:2012fst}%
  \BibitemOpen
  \bibfield  {author} {\bibinfo {author} {\bibfnamefont {M.}~\bibnamefont {Sharif}}\ and\ \bibinfo {author} {\bibfnamefont {S.}~\bibnamefont {Azeem}},\ }\bibfield  {title} {\bibinfo {title} {{Cosmological Evolution for Dark Energy Models in f(T) Gravity}},\ }\href {https://doi.org/10.1007/s10509-012-1172-9} {\bibfield  {journal} {\bibinfo  {journal} {Astrophys. Space Sci.}\ }\textbf {\bibinfo {volume} {342}},\ \bibinfo {pages} {521} (\bibinfo {year} {2012})},\ \Eprint {https://arxiv.org/abs/1305.0739} {arXiv:1305.0739 [gr-qc]} \BibitemShut {NoStop}%
\bibitem [{\citenamefont {RODRIGUES}\ \emph {et~al.}(2013)\citenamefont {RODRIGUES}, \citenamefont {HOUNDJO}, \citenamefont {MOMENI},\ and\ \citenamefont {MYRZAKULOV}}]{RODRIGUES_2013}%
  \BibitemOpen
  \bibfield  {author} {\bibinfo {author} {\bibfnamefont {M.~E.}\ \bibnamefont {RODRIGUES}}, \bibinfo {author} {\bibfnamefont {M.~J.~S.}\ \bibnamefont {HOUNDJO}}, \bibinfo {author} {\bibfnamefont {D.}~\bibnamefont {MOMENI}},\ and\ \bibinfo {author} {\bibfnamefont {R.}~\bibnamefont {MYRZAKULOV}},\ }\bibfield  {title} {\bibinfo {title} {Planar symmetry in f(t) gravity},\ }\href {https://doi.org/10.1142/s0218271813500430} {\bibfield  {journal} {\bibinfo  {journal} {International Journal of Modern Physics D}\ }\textbf {\bibinfo {volume} {22}},\ \bibinfo {pages} {1350043} (\bibinfo {year} {2013})}\BibitemShut {NoStop}%
\bibitem [{\citenamefont {Adak}\ \emph {et~al.}(2013)\citenamefont {Adak}, \citenamefont {Sert}, \citenamefont {Kalay},\ and\ \citenamefont {Sari}}]{Adak:2008gd}%
  \BibitemOpen
  \bibfield  {author} {\bibinfo {author} {\bibfnamefont {M.}~\bibnamefont {Adak}}, \bibinfo {author} {\bibfnamefont {O.}~\bibnamefont {Sert}}, \bibinfo {author} {\bibfnamefont {M.}~\bibnamefont {Kalay}},\ and\ \bibinfo {author} {\bibfnamefont {M.}~\bibnamefont {Sari}},\ }\bibfield  {title} {\bibinfo {title} {{Symmetric Teleparallel Gravity: Some exact solutions and spinor couplings}},\ }\href {https://doi.org/10.1142/S0217751X13501674} {\bibfield  {journal} {\bibinfo  {journal} {Int. J. Mod. Phys. A}\ }\textbf {\bibinfo {volume} {28}},\ \bibinfo {pages} {1350167} (\bibinfo {year} {2013})},\ \Eprint {https://arxiv.org/abs/0810.2388} {arXiv:0810.2388 [gr-qc]} \BibitemShut {NoStop}%
\bibitem [{\citenamefont {Stephani}\ \emph {et~al.}(2003)\citenamefont {Stephani}, \citenamefont {Kramer}, \citenamefont {MacCallum}, \citenamefont {Hoenselaers},\ and\ \citenamefont {Herlt}}]{Stephani:2003tm}%
  \BibitemOpen
  \bibfield  {author} {\bibinfo {author} {\bibfnamefont {H.}~\bibnamefont {Stephani}}, \bibinfo {author} {\bibfnamefont {D.}~\bibnamefont {Kramer}}, \bibinfo {author} {\bibfnamefont {M.~A.~H.}\ \bibnamefont {MacCallum}}, \bibinfo {author} {\bibfnamefont {C.}~\bibnamefont {Hoenselaers}},\ and\ \bibinfo {author} {\bibfnamefont {E.}~\bibnamefont {Herlt}},\ }\href {https://doi.org/10.1017/CBO9780511535185} {\emph {\bibinfo {title} {{Exact solutions of Einstein's field equations}}}},\ Cambridge Monographs on Mathematical Physics\ (\bibinfo  {publisher} {Cambridge Univ. Press},\ \bibinfo {address} {Cambridge},\ \bibinfo {year} {2003})\BibitemShut {NoStop}%
\end{thebibliography}%
\end{document}